\documentclass[aps,preprint,amssymb,12pt,floatfix]{revtex4}
\pdfoutput=1

\usepackage{subfigure}
\usepackage{graphicx}
\usepackage{lscape,graphicx}
\usepackage{rotating}
\usepackage{epstopdf}
\usepackage{color}
\usepackage{amsmath,amsthm}
\usepackage{wrapfig}

\begin{document}

\title
{Link between allosteric signal transduction and functional dynamics in a multi-subunit enzyme: \emph{S}-adenosylhomocysteine hydrolase}

\author{Yoonji Lee$^{1}$, Lak Shin Jeong$^{1,2}$, Sun Choi$^{1*}$, and Changbong Hyeon$^{3}$}
\thanks{To whom correspondence should be addressed : hyeoncb@kias.re.kr, sunchoi@ewha.ac.kr}
\affiliation{$^1$College of Pharmacy, Division of Life and Pharmaceutical Sciences, and National Core Research Center for Cell Signaling and Drug Discovery Research\\
$^2$Department of Bioinspired Sciences, Ewha Womans University, Seoul 120-750\\
$^3$School of Computational Sciences, Korea Institute for Advanced Study, Seoul 130-722, Korea
}

\begin{abstract}
\emph{S}-adenosylhomocysteine hydrolase (SAHH), a cellular enzyme that plays a key role in methylation reactions including those required for maturation of viral mRNA, is an important drug target in the discovery of antiviral agents. While targeting the active site is a straightforward strategy of enzyme inhibition, evidences of allosteric modulation of active site in many enzymes underscore the molecular origin of signal transduction. Information of co-evolving sequences in SAHH family and the key residues for functional dynamics that can be identified using native topology of the enzyme provide glimpses into how the allosteric signaling network, dispersed over the molecular structure, coordinates intra- and inter-subunit conformational dynamics. 
To study the link between the allosteric communication and functional dynamics of SAHHs, we performed Brownian dynamics simulations by building a coarse-grained model based on the holo and ligand-bound structures. 
The simulations of ligand-induced transition revealed that the signal of intra-subunit closure dynamics is transmitted to form inter-subunit contacts, which in turn invoke a precise alignment of active site, followed by the dimer-dimer rotation that compacts the whole tetrameric structure. 
Further analyses of SAHH dynamics associated with ligand binding 
provided evidence of both induced fit and population shift mechanisms, and also showed that the transition state ensemble is akin to the ligand-bound state. 
Besides the formation of enzyme-ligand contacts at the active site, the allosteric couplings from the residues distal to the active site is vital to the enzymatic function.
\end{abstract}
\maketitle
\section{Introduction}
\emph{S}-adenosylhomocysteine (SAH) hydrolase catalyzes the hydrolytic cleavage of SAH to adenosine and {\footnotesize L}-homocysteine. 
Inhibition of this enzyme causes the accumulation of SAH, and consequently suppresses \emph{S}-adenosyl-{\footnotesize L}-methionine dependent transmethylation via a feedback inhibition mechanism. 
Since the methylation at the 5'-terminus of mRNA is crucial for the viral replication, 
SAHH is a promising target for the discovery of broad spectrum antiviral agents \cite{de2002strategies}. 

Developing therapeutic agents that can directly bind and regulate the active site of a biological target has been a dominant pharmacological strategy \cite{kuntz1992Science,robert1997ChemRev}. 
For SAHH, various adenosine analogues, including carbocyclic adenosine, neplanocin A, 3-deazaneplanocin A \cite{de2002strategies}, and fluoro-neplanocin A (F-NpcA), are the recently developed inhibitors that directly target the active site \cite{lee2011JMC}. However, conformational flexibility of enzyme structures gleaned in x-ray, NMR experiments and the presence of allosteric site revealed in mutational studies highlight the allosteric couplings of residues distal to the active site as an another important principle in drug design strategy \cite{hardy2004COSB,goodey2008NCB,swain2006COSB}. 
Although the formation of specific enzyme-substrate contacts in the active site is required for the catalytic activity, allosteric orchestration among residues, dispersed over the molecular architecture, is also essential to regulate conformational fluctuations, so as to assist a precise positioning of catalytic elements in the active site \cite{knowles1991Nature}. 
SAHH, an enzyme consisting of chemically identical four subunits, each of which undergoes the open-to-closed (O$\rightarrow$C) transition in response to substrate binding, is an interesting system to study the link between enzymatic function and allosteric dynamics beyond monomeric enzyme \cite{schnell2004ARBBS,Chen07JMB,Taylor01ChemRev,Hyeon09PNAS,Kern07Nature,Whitford07JMB,kamata2004structure}.   

Structure, dynamics and catalytic function are the three main themes in understanding enzymes \cite{goodey2008NCB,schnell2004ARBBS,frauenfelder2001PNAS,Osborne2001BC,henzler2007Nature,hammes2006ARBC,bhabha2011Science}. 
For the past decades, much effort has been made to decode the link between the allosteric signaling of enzymes, conformational dynamics, and their function by using both theories \cite{Chen07JMB,Whitford07JMB,Lockless99Science,halabi2009cell,Zheng05Structure,balabin2009PNAS} and experiments \cite{Kern07Nature,bhabha2011Science,dhillon2002EMBOJ,Yang07PNAS,lee2008surface} 
, and has recently been extended to the studies of molecular motors \cite{Hyeon07PNAS,Hyeon07PNAS2,Chen10PNAS}. 
To gain microscopic understanding to the allostery in SAHH and its implication to the catalysis, we employed multifaceted computational approaches:  
(i) Statistical coupling analysis (SCA) \cite{Lockless99Science,Suel2002NSMB,Dima06ProtSci} was used to reveal networks of co-evolving amino acid residues in the SAHH family. 
(ii) Structural perturbation method (SPM) adapting normal mode analysis (NMA) identified the network of hot spot residues associated with the functional motion of SAHH \cite{Zheng05Structure}. 
(iii) While the SCA and SPM provide hints as to the correlation between allosteric signaling network and conformational dynamics, it is difficult to gain from these two static analyses further insights into a large scale conformational change, such as a shifting of statistical ensemble and heterogeneity in the transition routes. To this end, we performed Brownian dynamics (BD) simulations of the O$\rightarrow$C transition of tetrameric SAHH in response to ligand binding by using a structure-based coarse-grained model \cite{hyeon2011NatureComm,Hyeon09PNAS,Hyeon06Structure,Hyeon06PNAS,Chen10PNAS,Chen07JMB}. 

Among the various stages of the SAHH enzymatic cycle that involve multiple catalytic processes \cite{porter1993JBC,yang2003biochem}, which occur on time scales of $\sim(10-100)$ $ms$, main focus of our simulation is on the fast ($\sim(1-100)$ $\mu s$) conformational dynamics of SAHH before, after as well as in the process of substrate binding. 
Despite the large time scale gap between catalysis and conformational dynamics associated with ligand binding, 
it is important to understand the conformational fluctuations and dynamics since they lie at the core of allosteric regulation of catalytic power in enzymes. 
In this paper, by analyzing the dynamics resulting from our simulation, we highlight the link between the functional dynamics and allosteric signaling implicated in the SCA and SPM.

\section{Methods}

\begin{figure*}
\centering\includegraphics[width=6.2in]{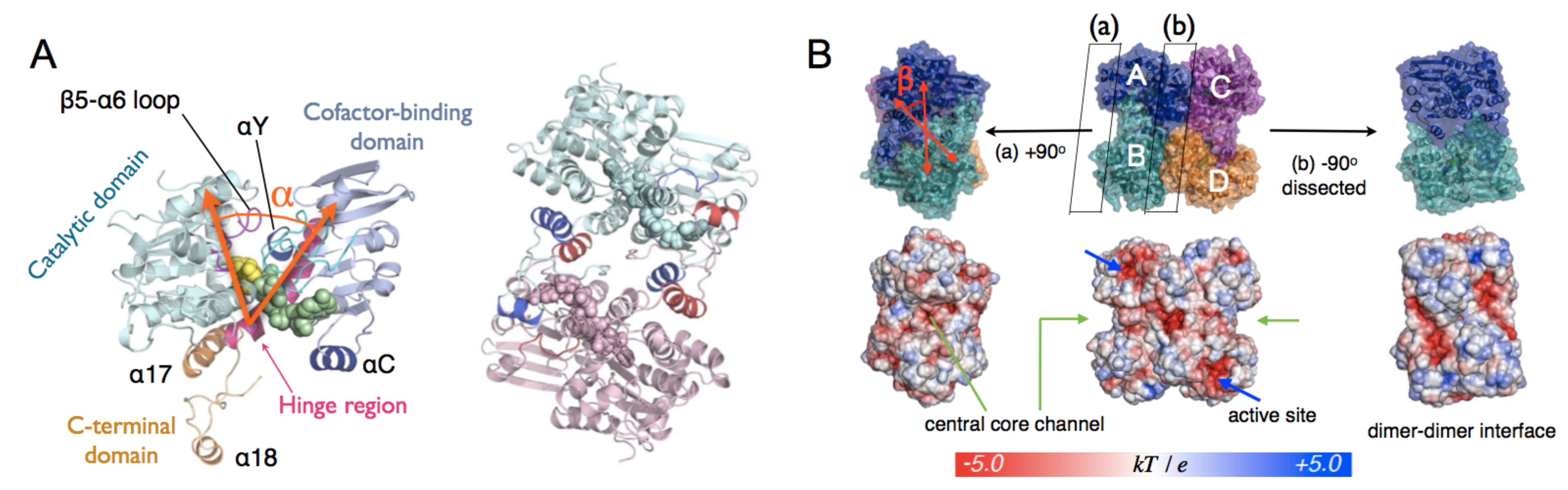}
 \caption{Structure of SAHH. 
{\bf A}. A subunit structure consisting of the catalytic (residues 1-181 and 355-385), cofactor-binding (residues 197-351) and C-terminal domains (residues 386-432) are shown on the left. The angle $\alpha$ is defined among W112, K186, and W310. 
A dimer structure assembled by exchanging the C-terminal domains is shown on the right with structural motifs associated with the dimer interactions being colored in blue and red.
{\bf B}. Homo-tetrameric structure. Each monomer is represented by blue, cyan, purple and orange surfaces.
$\beta$ is the angle between the two vectors, one  from CM$_\mathrm{A}$ to CM$_\mathrm{B}$ and the other from CM$_\mathrm{C}$ to CM$_\mathrm{D}$, where CM$_\mathrm{X}$ is the center of position of a subunit X. 
Electrostatic potentials on the molecular surface calculated by solving Poisson-Boltzmann equation at 0.2 M salt condition show negatively charged central core channel and active site. 
The side view of the structure and the dimer-dimer interface are shown on the left and right, respectively. 
\label{Fig_structure}}
 \end{figure*}

{\bf SAHH structure:}  
A subunit structure, consisting of 432 residues, retains catalytic, cofactor-binding and C-terminal domains. 
An alpha helix ($\alpha$A1) and loops (residues 182-196, 352-354) join the catalytic domain with the cofactor-binding domain, forming a hinge region (Fig.\ref{Fig_structure}A left, see also Figure S1). 
The interaction between C-terminal domain of a subunit and cofactor-binding domain of the partner subunit (Fig.\ref{Fig_structure}A right) is employed to assemble two dimer complexes (AB and CD) into the tetramer. 
A negatively charged central core channel constitutes the interface between A and B (or C and D) (Fig.\ref{Fig_structure}B). 
A further assembly between the two dimers (AB and CD), whose interface exhibits electrostatic complimentarity, forges an oblate-like tetrameric structure 
(Fig.\ref{Fig_structure}B).  
Depending on the presence of ligand in the binding cleft between catalytic and cofactor-binding domains, SAHHs select either open or closed conformation whose root-mean-square deviation (RMSD, $\Delta_{OC}$) is 4.2 \AA.  
The difference of the two subunit structures can also be quantified using an angle, $\alpha$, defined among W112, K186 and W310, which is $\alpha=70^o$ for the open and $\alpha=50^o$ for the closed structures (Fig.\ref{Fig_structure}A). 
In addition, an angle $\beta$ measures the relative orientation between AB and CD dimers; $\beta=3^o$ for the open and  $\beta=15^o$ for the closed structures (Fig.\ref{Fig_structure}B).\\

\noindent{\bf Statistical coupling analysis (SCA) on SAHH family:} 
If two amino acid residues in a protein are functionally or energetically coupled, a mutation at one site is expected to influence the other sites to level off the energetic perturbation, so that the protein can retain its functional state. 
While double or triple mutant experiment is laborious to prove this proposition, 
evolutionary data imprinted in multiple sequence alignment (MSA) of a protein family still allow one to infer the 
coupling between two distal sites. 
Statistical coupling analysis (SCA) is a method to identify a set of co-evolving amino acid sites by performing mutation ``experiment'' on the MSA of a protein family \cite{Lockless99Science,Suel2002NSMB,Dima06ProtSci}.

The whole sequence space of SAHH family, which includes 1156 sequences over the entire organisms, was clustered by their sequence identities, and 252 sequences under 75\% identities were finally selected. This diverse sequences of SAHH family were used to produce MSA using the multiple sequence comparison by log-expectation (MUSCLE) program \cite{edgar2004muscle}. 
Each of the sequences has 1,242 positions including gaps in the full MSA. 
Using multiple sequence alignment (MSA) of SAHH family, we calculated the statistical free energy like function defined as  
$\Delta G_i/k_BT^*=\sqrt{\tfrac{1}{C_i}\sum_{\alpha=1}^{20}\left[p_i^{\alpha}\log{(p_i^{\alpha}/p_{\alpha})}\right]^2}$ 
where $C_i$ is the number of types of amino acid at position $i$ along the sequence, $\alpha$ denotes amino acid species, $p_i^{\alpha}$ is the frequency of an amino acid $\alpha$ at position $i$, 
and $p_{\alpha}$ is the frequency of an amino acid $\alpha$ in the full MSA. $\Delta G_i$ (Fig.\ref{Fig_SCA}A) identifies highly conserved sequence positions in ligand- and cofactor-binding site and the C-terminal domain that constitutes a part of the cofactor-binding site (Fig.\ref{Fig_SCA}B). 

Further analysis of the MSA enables us to identify the set of mutually correlated residues in SAHH from an evolutionary perspective. The statistical coupling analysis (SCA) \cite{Lockless99Science,Suel2002NSMB,Dima06ProtSci} calculates  $\Delta\Delta G_{ij}/k_BT^*(=\sqrt{\tfrac{1}{C_i}\sum_{\alpha=1}^{20}\left[p_{i,R}^{\alpha}\log{(p_{i,R}^{\alpha}/p_{\alpha})}-p_i^{\alpha}\log{(p_{i}^{\alpha}/p_{\alpha})}\right]^2}$ where $p_{i,R}^{\alpha}$ is the frequency of amino acid in a restricted subalignment with respect to the position $j$), a score matrix that quantifies the effect of sequence "perturbation" at position $j$ on position $i$ by restricting the full MSA into a subalignment with the most conserved sequence at the $j$-th position and quantifying the sequence variation at the $i$-th position of subalignment with respect to the same position of the full MSA. 
Since the choice of appropriate size for subalignments ($f$) is critical in obtaining statistically significant correlations between residues in a protein family, the smallest value of $f$ should be chosen to satisfy the central limit theorem \cite{Dima06ProtSci}. In order to determine the suitable $f$ value, we build subalignments from the MSA with different $f$ ($<1$) values by randomly choosing $fN_{MSA}$ sequences, where $N_{MSA}$ is the size of the full MSA. For a given $f$, we generated 1000 subalignments and computed $\overline{\langle\Delta G_i^s\rangle_f}=\tfrac{1}{1000}\sum_{k=1}^{1000}\overline{\Delta G_i^s}$  and $\sigma_f^2=\overline{(\Delta G_i^s)^2}-\overline{\Delta G_i^s}^2$, where  $\overline{\Delta G_i^s}$ means the average $\Delta G_i$  values in a randomly chosen subalignment. By analyzing the distribution of these values, we selected $f=0.5$ to statistically significant subalignment size (Figure S2A), and 213 positions were allowed to be perturbed in this condition. In the $\Delta\Delta G_{ij}/k_BT^*$ scoring matrix, $\Delta\Delta G_{ij}$ with a high value signifies that the $i$-th position is susceptible to the perturbation at the $j$-th position. After filtering out the noise, the allosteric signaling network can be obtained by adopting the coupled two-way clustering algorithm to cluster the position and perturbation elements of  $\Delta\Delta G_{ij}$ exhibiting a high signal \cite{getz2000PNAS,Getz03Bioinformatics}. \\

\noindent{\bf Elastic network model (ENM) and structural perturbation method (SPM): } In ENM, a protein structure is represented as a network of beads connected with mechanical springs of uniform force constant $k_0$. 
The potential energy for ENM given as $E_{ENM}=\tfrac{1}{2}\sum_{ij}k_0(r_{ij}-r_{ij}^o)^2\Theta(r_{ij}^o-R_c)$, where $r_{ij}$ is the distance between residues $i$ and $j$, and $\Theta(\ldots)$ is the Heaviside step function, connects two coarse-grained center ($C_{\alpha}$ atoms) if the distance $r_{ij}^o$ in the native structure is within a cutoff distance $R_c$.  
By calculating the Hessian matrix of $E_{ENM}$ one can obtain normal modes ($\vec{\nu}_M$, $M\geq 7$, $1\leq M\leq 6$ are associated with translational and rotational degrees of freedom) of a protein architecture. Given the open ($\vec{r}_O$) and closed structures ($\vec{r}_C$) of SAHH, the overlap between $\vec{\nu}_M$ and the O$\rightarrow$C conformational change ($\vec{r}_{O\rightarrow C}$) is assessed by calculating $\cos{(\vec{\nu}_M\cdot \vec{r}_{O\rightarrow C})}$ (Fig.\ref{Fig_SPM}). 

The SPM assesses the importance of a residue in the elastic network by locally perturbing the residue and probing its response. The perturbation is realized by changing the force constant of the springs that link the residue and its neighbors. 
The response to the perturbation in the frequency of mode $M$, which overlaps the most with the O$\rightarrow$C structural transition, is calculated using $\delta \omega(M,n)=\vec{\nu}_M^T\cdot\delta H\cdot \vec{\nu}_M$  where $\delta H$ is the Hessian matrix of the following perturbed energy potential: $\delta E_{ENM}=\tfrac{1}{2}\sum_{ij}\delta k_0(r_{ij}-r_{ij}^o)^2\Theta(r_{ij}^o-R_c)$. The higher $\delta\omega(M,i)$ signifies an importance of $i$-th residue with respect to the mode $M$. \\

\noindent{\bf Energy potential and simulations: } To define the energy potential of SAHH, we used an x-ray crystal structure of human SAHH structure (PDB id: 3NJ4) that contains F-NpcA as the closed structure and a homology model based on rat SAHH as the open (or holo) structure \cite{lee2011JMC}. 
The open structure of SAHH should be stable in the absence of ligand, 
but the formation of the inter-residue contacts incorporated from the closed structure should bring SAHH from the open to closed forms in the presence of SAHH-ligand interactions, by further stabilizing the ligand-bound SAHH structure \cite{Hyeon09PNAS}. 
To implement this dynamical behavior, we used the self-organized polymer (SOP) energy potential \cite{Chen07JMB,Hyeon09PNAS,Hyeon06Structure}: $H_{tot}=H^{\mathrm{SAHH}}+H^{\mathrm{lig}}+H^{\mathrm{SAHH\cdot lig}}$ where $H^{\mathrm{SAHH}}=\sum_{i=\mathrm{A,B,C,D}}H_{\mathrm{intra}}^{\mathrm{SAHH}}(i)+\sum_{i>j}H_{\mathrm{inter}}^{\mathrm{SAHH}}(i,j)$ and  $H^{\mathrm{SAHH}}= H_{\mathrm{FENE}}+H_{\mathrm{nb}}$ consists of the terms involving intra-subunit connectivity using finite extensible nonlinear elastic potential ($H_{\mathrm{FENE}}$) and non-bonded potential term ($H_{\mathrm{nb}}$) that stabilizes either one of open or closed monomer structure depending on the presence of ligand at its binding cleft. 
The energy contributions from the $H_{\mathrm{nb}}$ are the sum of open native contacts ($H_{\mathrm{nb}}^{(O)}$) and pure closed native contacts  $H_{\mathrm{nb}}^{(C\cap O^c)}$ \cite{Hyeon09PNAS}. The enzyme-ligand interaction ($H^{\mathrm{SAHH\cdot lig}}$) should contribute to the SAHH compaction. 

We simulated the SAHH conformational dynamics under an overdamped condition by integrating the Brownian dynamics algorithm \cite{Hyeon09PNAS,McCammonJCP78}.  
The ligand binding dynamics were simulated by releasing four ligands at $\approx$ 20 \AA\ away from each binding cleft under a periodic and weak harmonic constraint ($k$=0.035 pN/nm) imposed every 25 $\mu s$, which was used to ensure the ligands to bind the binding cleft within our simulation time. 
The initial distance of the ligands from each binding pocket provides enough time to randomize the orientation of the ligand before it reaches the binding cleft (see also Movies S1 and S2). 
While the form of SOP potential and simulation strategy described above are essentially the same as those used in the previous study on protein kinase A \cite{Hyeon09PNAS}, the additional term that takes into account the inter-subunit interaction $H_{\mathrm{inter}}^{\mathrm{SAHH}}$ substantially enriches the resulting dynamics.

\section{Results and Discussion}

{\bf Allosteric signaling network inferred from co-evolving residues: } 
Employing SCA to the MSA of SAHH family, we identified two co-evolving clusters of residues (Fig.\ref{Fig_SCA}C and Table 1 for the list). 
It is noteworthy that since SAHHs function as a tetramer, the co-evolving clusters revealed from monomer sequence are indeed the consequence of allosteric communication between residues spread over the entire tetrameric architecture. Most of co-evolving residues identified by the SCA are not well-conserved over the SAHH family except for the residues in C-terminal domain and C79 that is in direct contact with the ligand (Figure S2C). (i) The cluster 1 shows that a number of residues in green clusters around the active site are susceptible to the perturbation of residues in yellow clusters. Given that the active site is the core region for SAHH function, it is not surprising that the residues susceptible to perturbations are distributed around the binding cleft. Moreover, the region of central core channel is also detected to be a part of the main allosteric signaling network, which includes C195, R196, and E197. In particular, C195 is known to be a key residue for the catalysis and the maintenance of the reduction potential at the 3'-position \cite{yuan1996JBC}. 
(ii) The cluster 2 includes two spatially disconnected sets of residues in the C-terminal domain (from F425 to Y432) and the residues (G122, P123, D125) in the catalytic domain that are located away from the active site. A mutation in "cyan" clusters perturbs the residues in "red" clusters. 
According to the mutation studies, K426 and Y430 in this region are the critical residues that affect the cofactor affinity and/or the assembly of tetrameric structure \cite{li2007BC,ault1994JBC}. Furthermore, although it was not detected in the SCA, a mutation of R49 in direct contact with D125 that belongs to the cluster 2, is reported to dramatically reduce SAHH activity \cite{vugrek2009HM}. \\

\noindent{\bf Network of hot spot residues responsible for the collective dynamics of SAHH: } 
Normal mode analysis (NMA) based on the elastic network model (ENM) is used to grasp the caricature of SAHH dynamics evolving from the native topology.  
Along with the previous studies,\cite{wang2005domain,hu2008Prot} the covariance matrices calculated for the open and closed SAHH provide a general view of how the enzyme fluctuations in each form (see Figure 3A). 
The catalytic domain is more mobile than the cofactor-binding domain, and the overall flexibility decreases in the closed conformation. They also indicate that the cofactor-binding domain and the C-terminal domain of partner subunit are cross-correlated (see black arrows), in consistent with the structural organization of SAHH dimer where the C-terminal domains are exchanged between the adjacent subunits. Interestingly, the catalytic domain of A (or B) subunit is cross-correlated with that of D (or C) subunit (see red arrows).
The largest overlap of normal mode with the O$\rightarrow$C conformational change ($\vec{r}_{O\rightarrow C}$) (see Methods) is found with the lowest frequency normal mode (mode 7) in both the open and closed forms (Figures 3B left and 3C left). 
The closure motion of catalytic domain relative to the cofactor-binding domain and the relative rotation between AB and CD dimers are the main fluctuations characterizing the global dynamics of open and closed forms of SAHH.  

To determine "hot spot" residues controlling the O$\rightarrow$C conformational dynamics, we employed the structural perturbation method (SPM) \cite{Zheng05Structure}, which complements the results of SCA. The SPM mimics the point mutation of an enzyme by increasing the spring constant of a residue in the elastic network. Perturbation to a residue important for functional motion leads to a large change in the normal mode frequency (high-$\delta\omega$, see Methods). For the closure dynamics described by mode 7 of the open SAHH structure, key residues with high-$\delta\omega$  are mainly distributed in the hinge region (Fig.\ref{Fig_SPM}B right). 
Notably, this finding conforms to the recent time-resolved fluorescence anisotropy measurements that probed the domain motions of mutant SAHH, where importance of residues M351, H353, P354 at the hinge region was highlighted.\cite{wang2006Biochemistry}
On the other hand, for the dimer-dimer rotation dynamics described by the mode 7 of the closed structure, high-$\delta\omega$  residues are found at the interfaces between subunits (Fig.\ref{Fig_SPM}C right). Key residues controlling the next higher frequency modes are distributed mainly at the interfaces between the subunits, contributing to the concerted intersubunit dynamics (see Figure S3 for mode 9). \\

\noindent{\bf The global motion of SAHH during the O$\rightarrow$C transitions upon ligand binding: } 
The conformational dynamics and its dynamical response to ligand binding are prerequisite to understanding the enzyme function. To this end, we performed Brownian dynamics (BD) simulations using a coarse-grained model of the whole tetrameric structures in the holo and ligand-bound states (see Methods). Each trajectory consists of 750 $\mu s$ simulations; the first 250 $\mu s$ simulates the equilibrium stage of SAHH in the absence of ligand, and the next 500 $\mu s$ is for the O$\rightarrow$C transition dynamics in response to the ligand binding. We probed the ligand binding by using the distance between the fluorine atom of F-NpcA and C4 atom of the nicotinamide ring of the cofactor NADH (Figure S4). Among the total 60 simulated trajectories, we observed complete bindings of four ligands in 39 trajectories during our simulation time. A comparison of the fraction of native contact map ($Q_{ij}$) before and after the ligand binding revealed that $Q_{ij}$  increases and decreases depending on residue pairs (Fig.\ref{Fig_dynamics}A), suggesting that the structural reorganization due to $O\rightarrow C$ transition involves formation and breakage of multiple residue pairs (Fig.\ref{Fig_dynamics}A, 4B, and 4C).  
The $Q_{ij}$ between the $\beta$5-$\alpha$6 loop and $\alpha$Y helix (marked with blue spheres in Fig.\ref{Fig_dynamics}B), located at the entrance of the binding cleft, dramatically increases after the ligand binding, while the $Q_{ij}$ between $\alpha A1$, $\alpha 17$ and $\alpha 15$ (marked with red spheres in Fig.\ref{Fig_dynamics}B) decreases. 
In case of the inter-subunit contacts, key increases of residue contacts are made between the hinge region and $\alpha$C helix of the partner subunit (Fig.\ref{Fig_dynamics}C).                                                                                                                                                                                                                                                                                                                                                                                                                                                                                                                                                                                                                                                                                                                                                                                                                                                                                                                                                                                                                                                                                                                                                                                                                                                                                                                                                                                                                                                                                                                                                                                                                                                                                                                                                                                                                                                                                                                                                                                                                                                                                                                                                                                                                                                                                                                                                                                                                                                                                                                                                                                                                                                                                                                                                                                                                                                                                                                                                                                                                                                                                                                                                                                                                                                                                                                                                                                                                                                                                                                                                                                                                                                                                                                                                                                                                                                                                                                                                                                                                                                                                                                                                                                                                                                                                                                                                                    

Prior to the ligand binding, the enzyme structure remains largely in the open state as expected. 
Nevertheless, the fluctuation of the RMSD relative to the closed structure ($\Delta_C$) occasionally brings SAHH structure to $\Delta_C <2.5$ \AA\ even without ligand in the binding cleft (Fig.\ref{Fig_popul}B). 
As a result of  the ligand binding, the $\Delta_C$ decays from $3.7$ \AA\ to $2.9$ \AA\ in 34 $\mu s$ (Fig.\ref{Fig_dynamics}D left), the angle $\alpha$ decreases from $69^o$ to $53^o$ (Fig.\ref{Fig_dynamics}D middle), and the angle $\beta$ changes from $5^o$ to $\approx 12^o$ (Fig.\ref{Fig_dynamics}D right). \\

\noindent{\bf Kinetic hierarchy of the O$\rightarrow$C transition:} 
The inter-residue distance between $i$-th and $j$-th residues, $\langle d_{ij}(t)\rangle$, where $\langle\ldots\rangle$ denotes an ensemble average over trajectories, probes microscopic details of O$\rightarrow$C transitions described above. Histograms of $d_{ij}$, $P(d_{ij})$, before and after the ligand binding quantify the change in average and fluctuation of inter-residue distance (Fig.\ref{Fig_dynamics}). 
The average transition time for each residue pair, obtained from multi (mostly single) exponential fit to the $\langle d_{ij}(t)\rangle$, reveals the order of events that occur during the O$\rightarrow$C transition dynamics upon ligand binding (Fig.\ref{Fig_summary}). 

{\it Closure dynamics of the binding cleft}: The structural motifs in the binding cleft, composed of several loops and an alpha helix, directly interact with a ligand that mediates the association between the catalytic and cofactor-binding domains (Fig.\ref{Fig_structure}A). We employed several residue pairs in these loops as reporters of intra-subunit closure dynamics. The dynamics of residue pairs associated with G277, H301, A340 and R343 from three representative loops consisting of binding cleft show single exponential decays with time constants of $\tau=35-38$ $\mu s$ (Fig.\ref{Fig_hierarchy}A green panel). The ligand binding reduces not only the average distances between the catalytic and cofactor-binding domains but also their fluctuations (see the inset of green panel in Fig.\ref{Fig_hierarchy}A for D131-H301 pair). 
These results are in good agreement with the experimental data of fluorescence anisotropy measurements,\cite{yin2000Biochem} confirming that the intra-subunit closure dynamics within SAHH are substantially reduced after the ligand binding.
Among the residue pairs we probed, H301 is known to form key hydrogen bonding with the ligand \cite{lee2011JMC,wang2007BMCL}. Interestingly, some of residue pairs associated with L347, G348 and M351, located at the entrance of binding cleft, have to expand during the ligand binding to accommodate the ligand to the active site (Fig.\ref{Fig_hierarchy}A blue panel), reminiscent of partial unfolding and refolding observed in protein-protein or protein-ligand bindings \cite{Hyeon09PNAS,Miyashita03PNAS,schrank2009PNAS}. 

{\it Inter-subunit dynamics probed with the residue pairs at the subunit interface}: 
The dynamics of inter-subunit residue pairs are responsible for the concerted motion of tetrameric SAHH.  
The inter-subunit residue pairs, including K166-G418, F189-Q251 at AB (or CD) interface, and W17-I321, W17-K322, K20- I321 at AC (or BD) interface, exhibit a single exponential decay slower than that of the intra-subunit residue pairs (the magenta and cyan panels of Fig.\ref{Fig_hierarchy}A). 
It is of particular note that many of individual time traces of inter-subunit residue pairs show bimodal hopping transition (Fig.\ref{Fig_hierarchy}B); the histogram of  $d_{\mathrm{AC(BD):W17-I321}}$ can be described using a double gaussian function, centered at 8 \AA\ and 14 \AA\ in the absence of ligand. Upon ligand binding the population shifts toward  $d_{\mathrm{AC(BD):W17-I321}}\approx$ 8 \AA, reminiscent of population shift mechanism. 

{\it Two step mechanism of open-to-closed transition dynamics}: Comparison of transition times ($\tau$) of various residue pairs revealed an order of events that occur during the O$\rightarrow$C dynamics. 
The hierarchical order of $\tau$ (Fig.\ref{Fig_hierarchy}) suggests that transition dynamics progresses from intra-subunit closure dynamics to the inter-subunit contact formations.
Upon the ligand binding, (i) the intra-residue pairs in the cleft regions participating in the closure dynamics occurs with time scales of  $\tau\approx 33-38$ $\mu s$. Many of intra-residue pairs, located in particular at the entrance of the binding cleft (blue spheres in Fig.\ref{Fig_hierarchy}A) show nonmonotonic kinetics with transition times $\tau\approx 33-36$ $\mu s$, followed by $\alpha$ angle transition in 37 $\mu s$. (ii) The inter-subunit residue pairs (cyan and magenta spheres in Fig.\ref{Fig_hierarchy}A) and the intra-residue pairs at the central region of binding cleft (green spheres) decrease with decay times of  $\tau\approx 35-38$ $\mu s$. 
Of particular note is that the contact formation of residue pairs at the AC (or BD) interface precedes the full ordering of the binding cleft.    
Finally, the dimer-dimer rotation between AB and CD ($\beta\approx 4^o\rightarrow 11^o$) is completed in 43 $\mu s$. 

To summarize (see Figures 5 and 6), in response to the ligand binding, the closure dynamics of the binding cleft due to the formation of intra-subunit contacts (blue), probed with $\langle\alpha\rangle$, precedes both the inter-subunit residue contacts between AC or BD interface (cyan) and the more precise positioning of the active site (green). 
The rearrangement of interfacial contacts (magenta) between A and B (or C and D) subunits  and dimer-dimer rotation, measured with $\langle\beta\rangle$, occurs at the stage later than the closure dynamics measured with $\alpha$. 
From local compaction at the binding cleft to the global compaction of entire structure exists a hierarchical order in $O\rightarrow C$ transition dynamics. \\

\noindent{\bf Transition dynamics of SAHH in response to the ligand binding occurs via both modulation of energy landscape and selection of conformational ensemble:} 
Two limiting mechanisms, induced fit and population shift (or conformational selection) (Fig.\ref{Fig_popul}A), are often contrasted in literatures to account for the allosteric dynamics associated with ligand binding or protein-protein association
\cite{Hammes09PNAS,Itoh10PNAS}. 
In the SAHH dynamics, it is clear that the interaction between the ligand and SAHH ($H^{\mathrm{SAHH-lig}}$) is the main cause of the $O\rightarrow C$ transition. 
However, it is not clear whether the ligand binding selects a small subpopulation of the closed structure from a pre-existing ensemble or modulates the folding landscapes.  
From the perspective of statistical thermodynamics, this question may be answered by projecting the population of accessible conformational states on a map in reduced dimension and comparing the two populations obtained from the holo and ligand-bound conditions. 
 If the overlap between the statistical ensembles corresponding to the closed structure under the energy hamiltonians of the holo and ligand-bound states is substantial, then population shift is more plausible. 
In contrast, if there is little overlap between the statistical ensembles, then induced fit mechanism is favored.


Because of the attractive contact pairs incorporated from the closed structure (see Methods), SAHH structure in the open form can transiently visit compact (or closed-like) structures even in the absence of ligand (Fig.\ref{Fig_popul}B) \cite{bahar2007COSB}. We assessed the similarity between the ensemble of closed-like structures (CL) that satisfies $\Delta_C<R_c$, where $R_c=2.5$ \AA\ in holo-structure and the ensemble of closed structures (C) that also satisfies $\Delta_C<R_c$ in the presence of ligand (Figures 7C and 7D). 
Following the line of argument made above, if the enzyme recognizes the ligand solely by selecting a pre-existing conformational ensemble, a distribution of scattered plot using inter-residue distances $(d_{\mathrm{K103-G346}}, d_{\mathrm{D131-H301}})$ or angles $(\alpha,\beta)$  as surrogate reaction coordinates will show a substantial overlap of CL ensemble with C ensemble. 
On the contrary, under the induced fit mechanism, we expect a difference between the distributions of CL and C. 
In Figures 7C and 7D, C and CL structural ensembles show clear distinction, but with a certain degree of overlap, giving evidence for both induced fit and population shift mechanisms. 

Depending on the molecule being studied, the concentration of ligand, and the position of probe being attached, one of the two mechanisms becomes more dominant \cite{Hammes09PNAS}. 
For SAHH, the feature of conformational selection to the closed structure is observed in the residue pairs at the inter-subunit interface (see the dynamics of $d_{AC(BD):W17-I321}$ in \ref{Fig_hierarchy}B where one of the bimodal basins corresponding to the closed structure at $d_{AC(BD):W17-I321}\approx 8$ \AA\ becomes more populated upon the ligand binding) while the modulation of energy landscape, which alters the positions of energy minima, is more dominant in the dynamics of intra-subunit residue pairs.  \\

\noindent{\bf Characteristics of transition state ensemble of SAHH during the O$\rightarrow$C transition:}
Identifying characteristics of the transition state of an enzyme along the reaction coordinate associated with its functional motion is an important task in designing an inhibitor drug that can fit to the active site. To this end, we analyzed the transition state ensembles (TSEs) of SAHH conformations by using $\Delta_C(t)$ and $\Delta_O(t)$  as surrogate reaction coordinates \cite{Chen07JMB,Hyeon06PNAS}. Upon ligand binding, $\langle\Delta_C(t)\rangle$  increases and $\langle\Delta_O(t)\rangle$  decreases, 
and they cross at a certain moment, i.e., $\langle\Delta_O(t)\rangle=\langle\Delta_C(t)\rangle$  (\ref{Fig_dynamics}D left). 
Transition states can be collected from the individual time traces by imposing a condition $|\Delta_O(t)-\Delta_C(t)|<\delta$ with $\delta=0.01$ \AA\ for intra-subunit and $\delta=0.3$ \AA\ for inter-subunit dynamics (\ref{Fig_TSE}A). 

The TSE characterized by the pair of intra-subunit residue distances ($d^{TS}_{H55-L347}$, $d^{TS}_{G132-H301}$) or angles ($\alpha$,$\beta$) has a broad and heterogeneous distribution (\ref{Fig_TSE}B), implying that the transition routes are formed on a plastic energy landscape. 
A Tanford $\beta$-like paramter, defined using the equilibrated distances of open form ($x_O$) and closed form ($x_C$), $\beta_T=(x_{TS}-x_C)/(x_O-x_C)$, evaluates the similarity of TSE to the ligand-bound state \cite{Hyeon06PNAS}. We obtained $\beta_T=0.33$ and $0.44$ for the dynamics of both distance and angle pairs, suggesting that the TSEs resemble the ligand-bound (closed) state ($\beta_T<0.5$). 

\section{Concluding remarks}
\begin{figure}
\centering\includegraphics[width=7.0in]{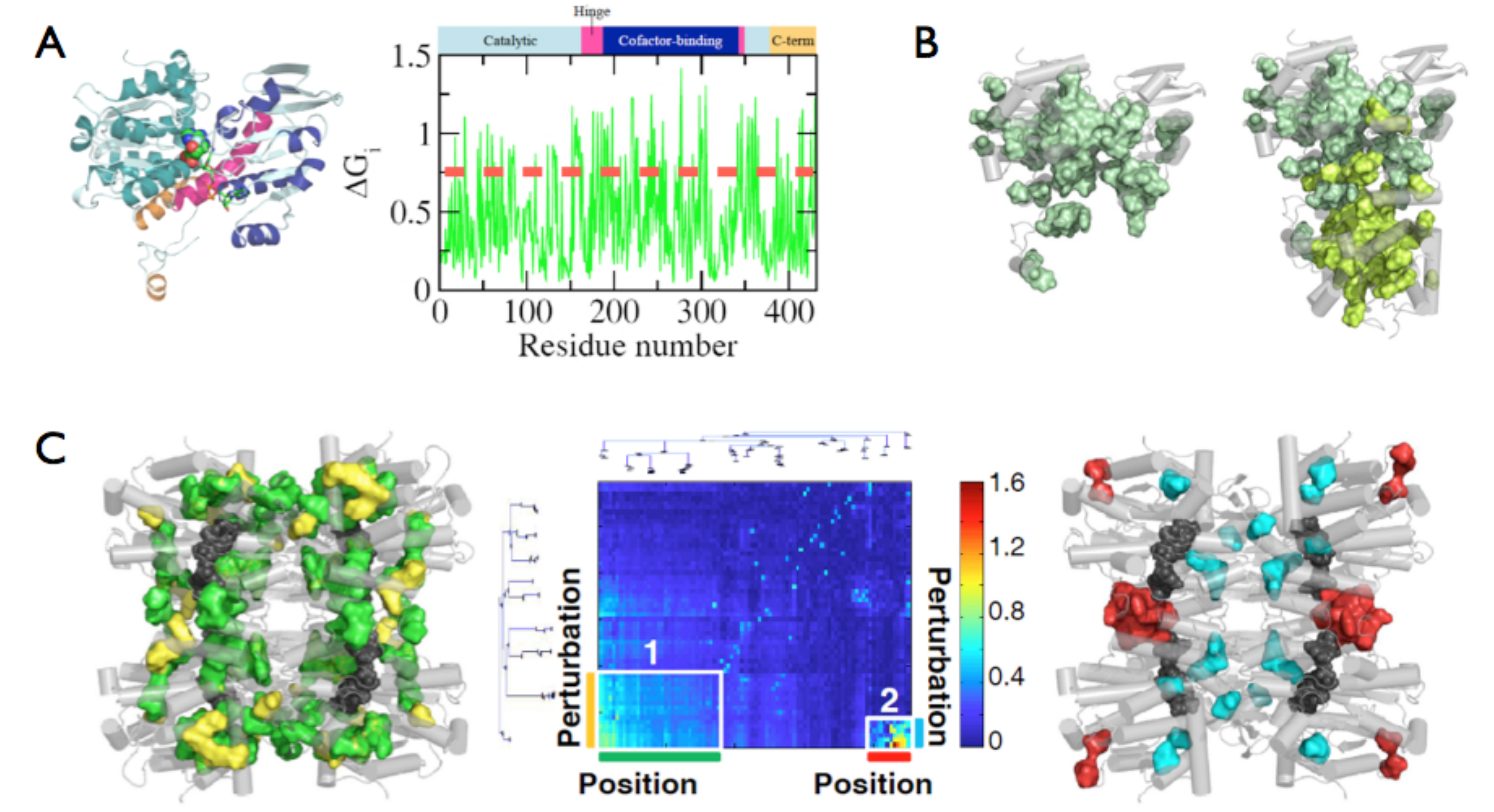}
 \caption{Statistical coupling analysis of SAHHs.
 {\bf A}. $\Delta G_i/kT^*$ value of the $i$-th residue in SAHH monomer. The catalytic (cyan), hinge (magenta), cofactor-binding (dark blue), and C-terminal tail domains are depicted in different colors. 
 {\bf B}. Highly conserved residues with $\Delta G_i>0.75$ from {\bf A} are mapped on the monomer (left) and dimer (right) structures.   
{\bf C}. Reordered $\mathrm{\Delta\Delta G_{ij}}$ matrix that identifies the two clusters (white boxes) is shown in the middle and the corresponding  two co-evolving clusters of sequences are depicted on the SAHH structure (see \emph{SI}). 
For cluster1, perturbation and position clusters are colored in yellow and green, respectively. For cluster2, perturbation and position clusters are colored in cyan and red, respectively. 
\label{Fig_SCA}}
 \end{figure}

\begin{figure}
\centering{\includegraphics[width=6.5in]{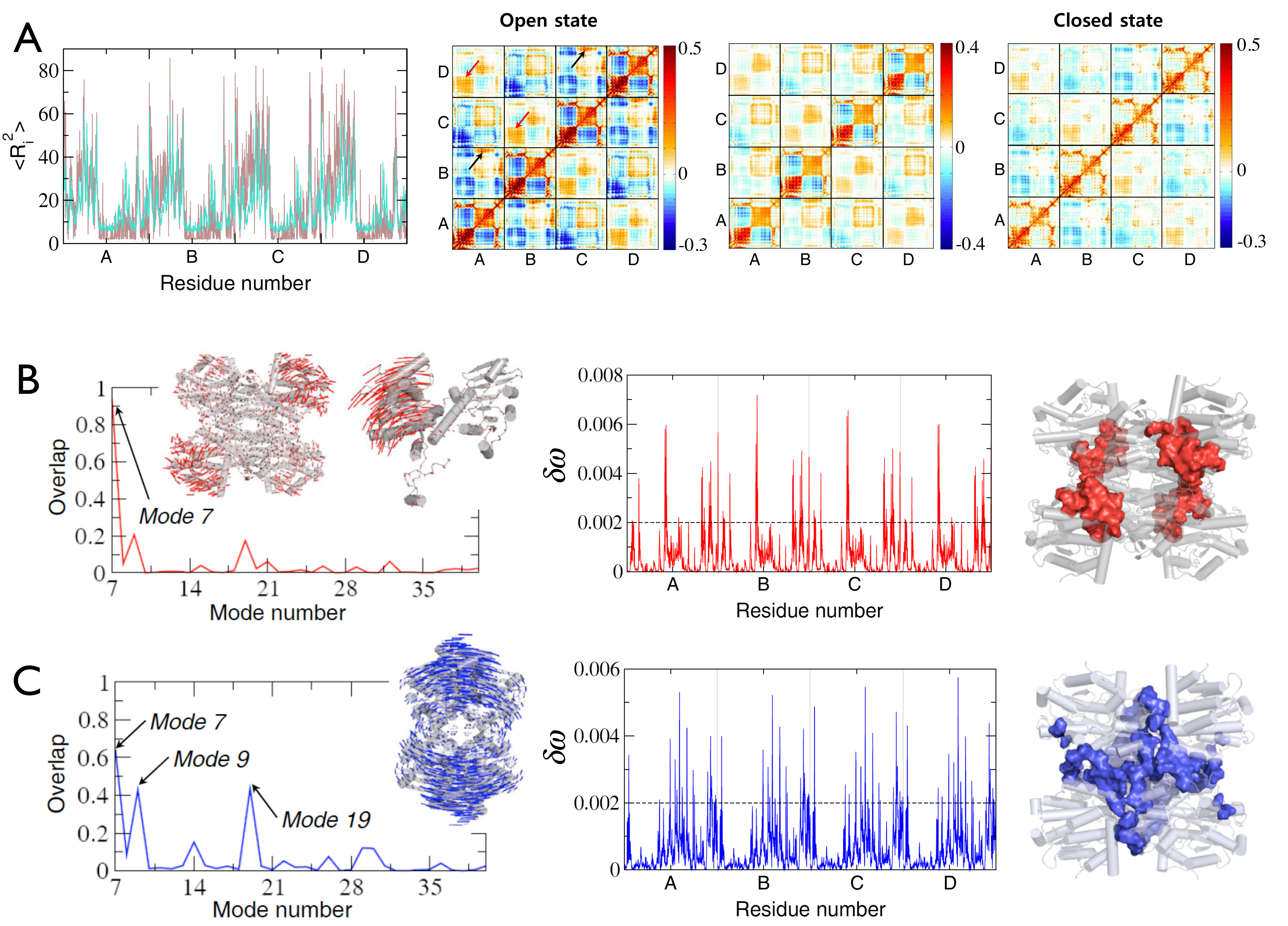}}
 \caption{Normal mode analysis and structural perturbation method applied to SAHH.
 {\bf A}. (Left) Comparison between mean square fluctuation (cyan) and crystallographic b-factor (brown) of the open form of rat SAHH (PDB code: 1B3R). 
(Right) Covariance matrices of SAHH open (left) and closed (right) forms. The differences between the two matrices are shown in the middle. Our calculations using ENM compare well with those in previous study \cite{wang2005domain}.
 {\bf B}. Overlap between $\vec{r}_{OC}$ and normal mode $M$ ($\vec{\nu}_M$) of the open structure. 
 Maximally overlapping normal mode ($M=7$) are displayed using lines in the structures (left). 
The effect of structural perturbation, $\delta \omega(M,i)$, is calculated for $\vec{\nu}_{M=7}$ (middle).  
The residues with high-$\delta\omega$ value with respect to $\vec{\nu}_{M=7}$ are depicted in the structure on the right and listed in the Table S1. 
{\bf C}. The same calculation for the closed structure by using $\vec{\nu}_{M=7}$ as the maximally overlapping mode with $\vec{r}_{OC}$. 
\label{Fig_SPM}}
 \end{figure}

\begin{figure*}
\centering{\includegraphics[width=4.8in]{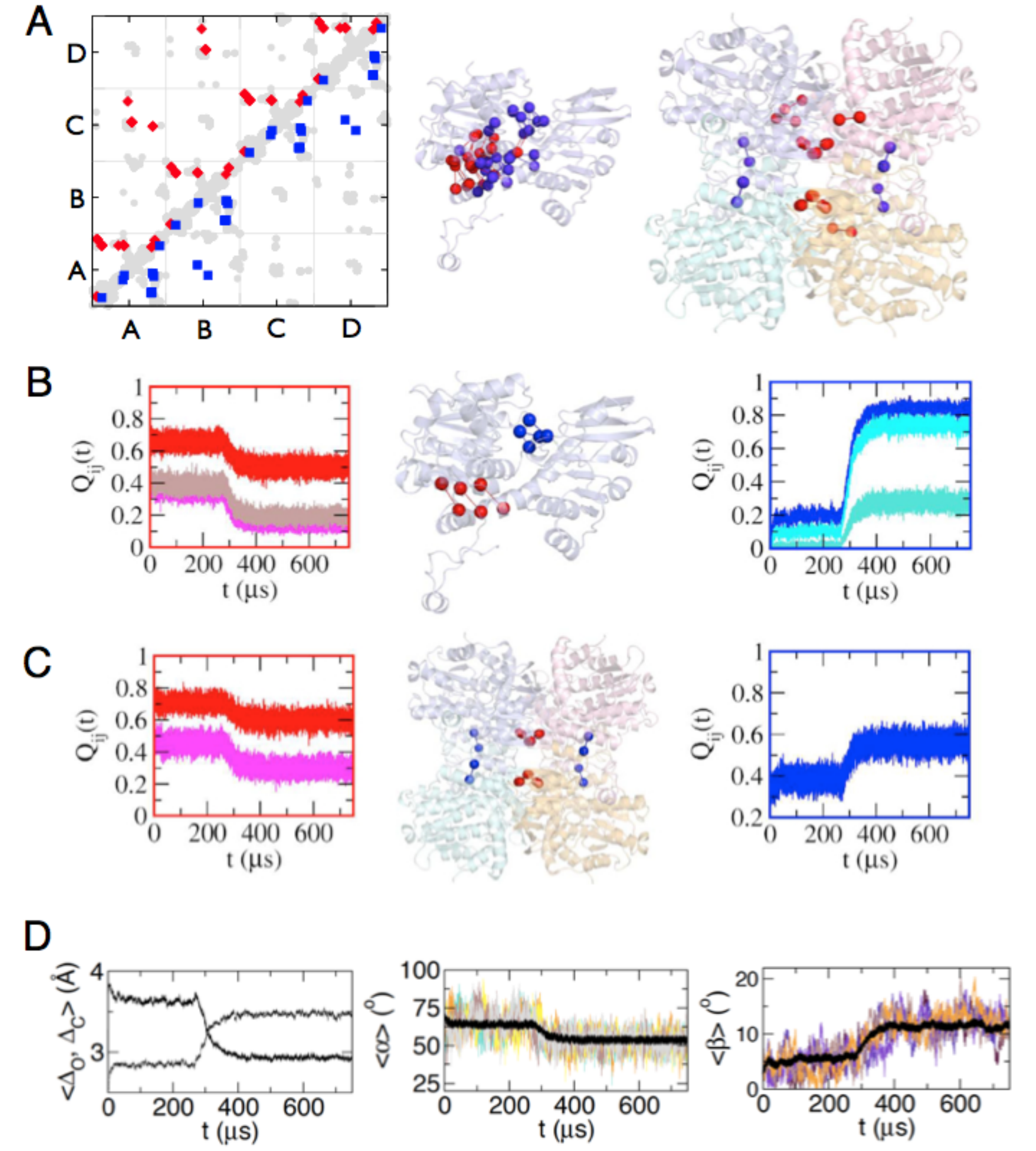}}
 \caption{BD simulations of the O$\rightarrow$C transitions in SAHH. 
{\bf A}. Native contact probability ($Q_{ij}$) changes as a result of the conformational change in response to the ligand binding. Among the native contact pairs, $Q_{ij}$ are colored in blue (red) when $Q_{ij}$ increases (decreases) more than $10$ \%. The position of corresponding residue pairs in a SAHH subunit  and at the subunit interface are depicted in the structures on the right. 
{\bf B, C}. Representative time traces of native contacts $Q_{ij}(t)$ for the intra- ({\bf B}) and inter-subunit ({\bf C}) residue pairs. The residue pairs with $\delta Q_{ij}(t)<0$ after ligand binding are marked in red, and those with $\delta Q_{ij}(t)>0$ are in blue. 189-361 (red), 185-364 (magenta), 367-394 (brown), 83-347 (blue), 83-348 (cyan), 82--343 (turquoise); AC(BD):231-235 (red), AC(BD):234-235 (magenta), AB(CD):185-247 (blue). 
{\bf D}. Time-dependent changes of root mean square deviation (RMSD) relative to the open and closed forms ($\langle\Delta_O\rangle$ and $\langle\Delta_C\rangle$), $\langle\alpha\rangle$ and $\langle\beta\rangle$ are shown from left to right.  
\label{Fig_dynamics}}
 \end{figure*}

\begin{figure*}
\centering{\includegraphics[width=5.5in]{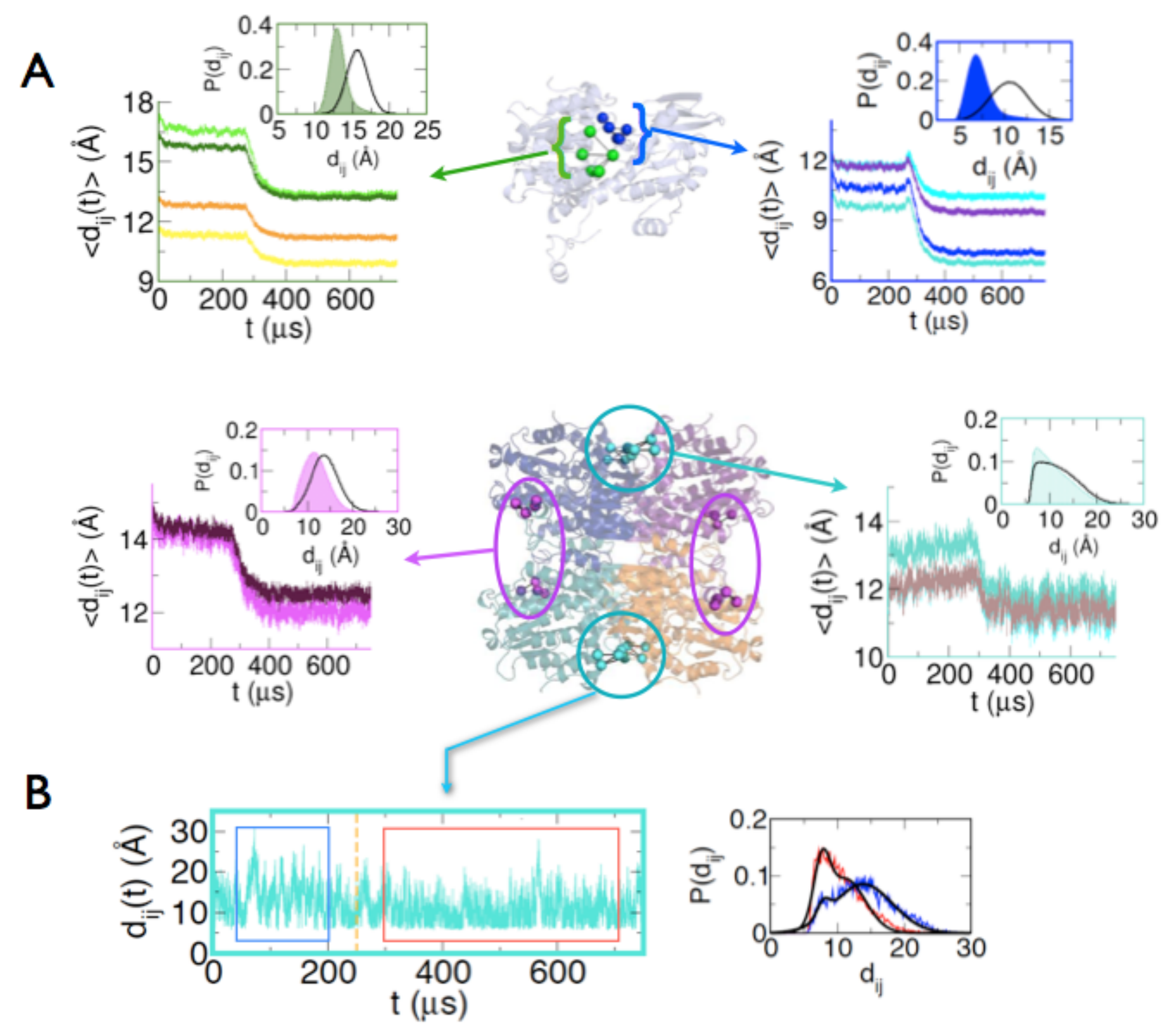}}
 \caption{Dynamics of intra- and inter-subunit residue pairs.  
{\bf A}. Intra- and inter-subunit dynamics averaged over the ensemble of trajectories probed by $\langle d_{ij}\rangle$ and the corresponding histogram, $P(d_{ij})$, before (black line) and after (colored area) the ligand binding.  
The intra-subunit residue pairs consisting of the catalytic site exhibit monotonic (green spheres) and non-monotonic decays (blue spheres).  
The inter-subunit residue pairs at AB (or CD) and AC (or BD) interfaces are displayed in purple and cyan spheres, respectively.  
The list of residue pair dynamics depicted in each panel is: 79-301, 131-301$^*$, 157-301, and 159-301 for green spheres; 83-347, 83-348$^*$ 83-351, and 85-351 for blue spheres; AB(CD) 166-418$^*$ and AB(CD) 162-416 for magenta spheres; AC(BD) 17-321$^*$, 17-322 and 20-321 for cyan spheres, where anterisk ($^*$) denotes the residue pairs whose histogram is shown in the insets.
{\bf B}. An individual time trace of $d_{\mathrm{AC(BD):W17-I321}}(t)$ at the AC subunit interface displays a biomodal hopping transition, and its population change is shown in the upper right panel. 
The histograms before and after the ligand binding (blue and red lines, respectively) fitted to a double Gaussian function (black lines) show that there is a ``shifting of population'' from $d_{\mathrm{AC(BD):W17-I321}}\approx 14$ \AA\ to 8 \AA.
\label{Fig_hierarchy}}
 \end{figure*}

\begin{figure*}
\centering{\includegraphics[width=5.8in]{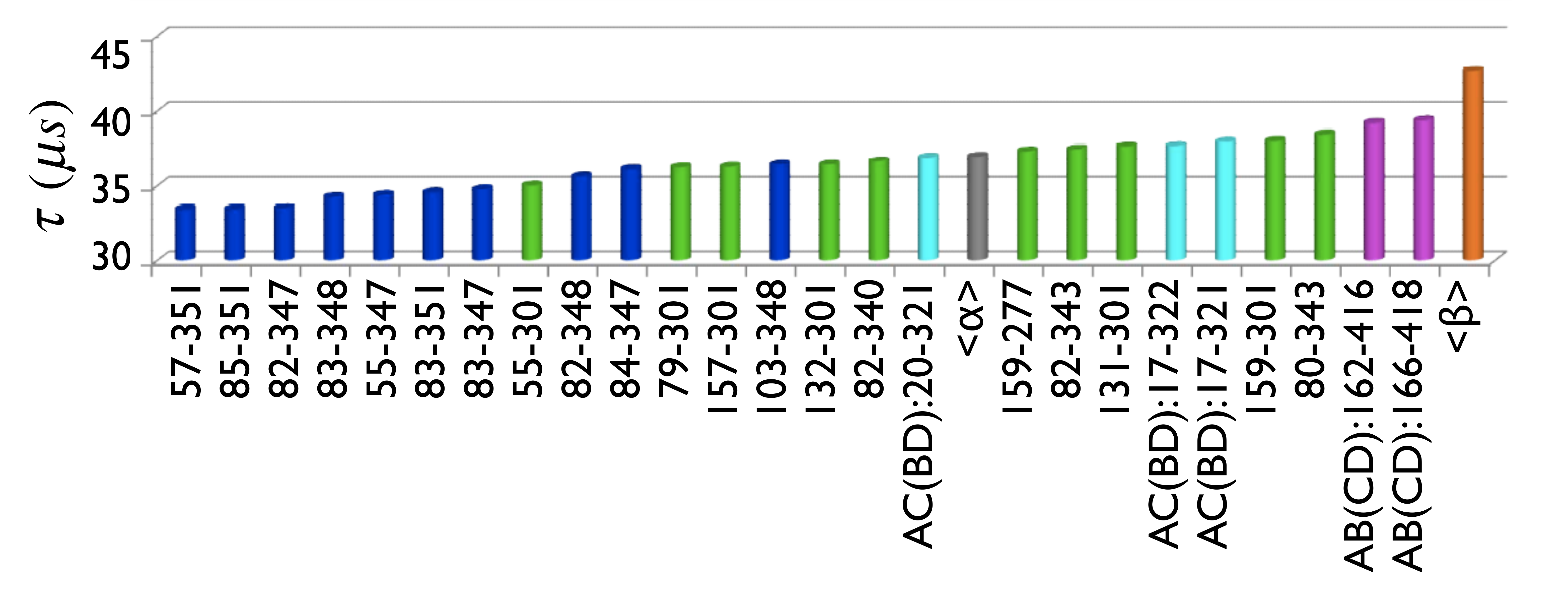}}
 \caption{Kinetic hierarchy revealed in the SAHH dynamics.  
Average transition times for various residue pairs discussed in \ref{Fig_hierarchy} and $\alpha$, $\beta$ angle dynamics are shown in an increasing order. The colors of bar graphs are identical to those used in the previous figures. 
\label{Fig_summary}}
 \end{figure*}

\begin{figure}
\centering{\includegraphics[width=5.2in]{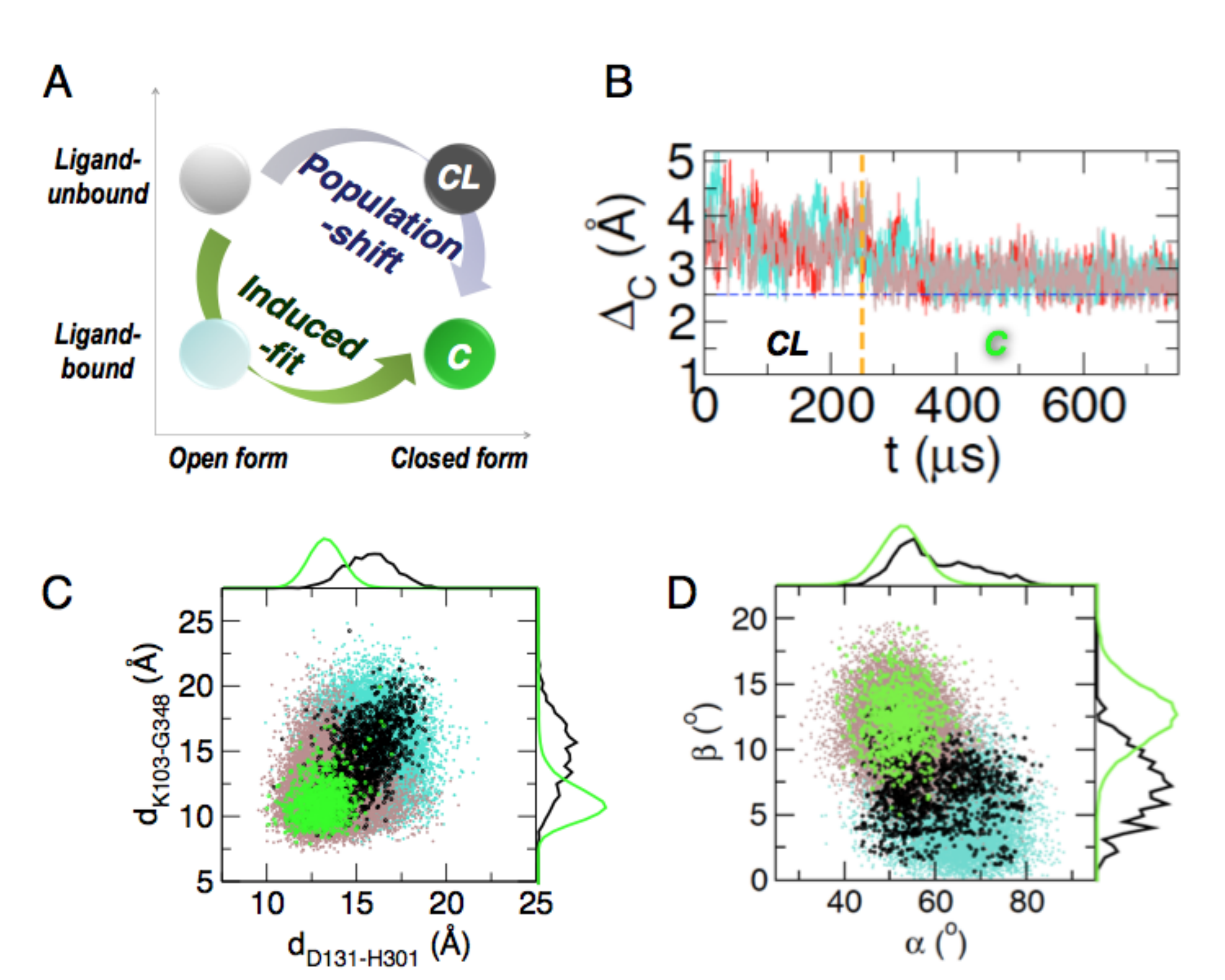}}
 \caption{Comparison between the closed-like (CL) and closed structures (C). 
 {\bf A}. A schematic diagram of molecular recognition processes contrasting induced fit and conformational selection mechanisms.
 {\bf B}. Exemplary time traces of $\Delta_C$ visiting the CL state prior to the ligand binding ($t<250$ $\mu sec$, $\Delta_C<R_c=2.5$ \AA) and the C state in the presence of ligand ($t>250$ $\mu sec$, $\Delta_C<R_c=2.5$ \AA).  
 {\bf C}. Scattered plot using two intra-subunit residue distances. Ensembles of CL and C states are displayed in black and green, respectively, and their histograms are shown in the upper and right side of the graphs. For comparison, the projected space sampled by the whole trajectories before and after the ligand binding are shown in cyan and brown.
 {\bf D}. Scattered plot of CL and C ensembles using $\alpha$ and $\beta$. 
 \label{Fig_popul}}
 \end{figure}

\begin{figure}
\centering{\includegraphics[width=5.2in]{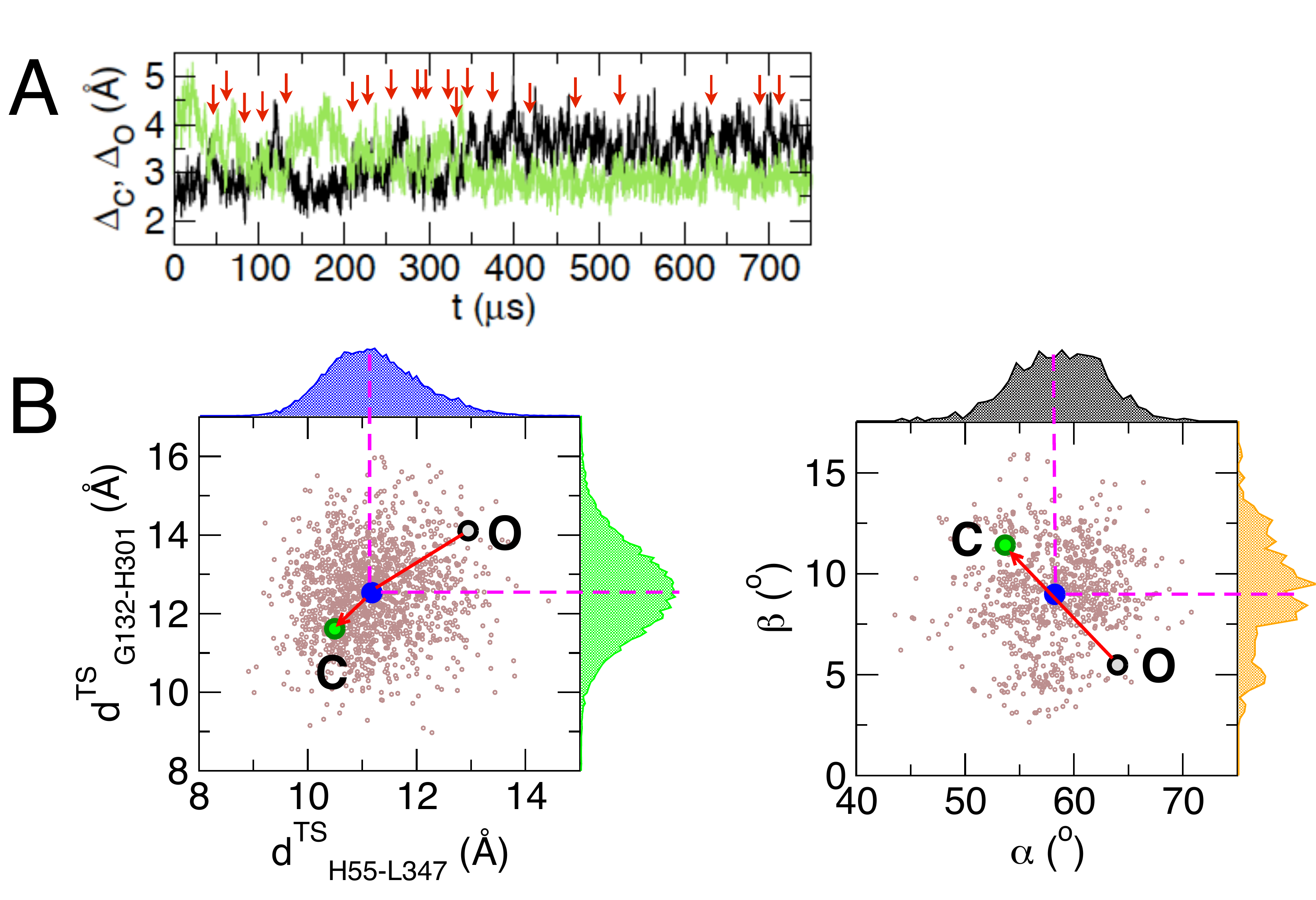}}
 \caption{Transition state ensembles of the O$\rightarrow$C transitions.  
 {\bf A}. Time-dependent
changes of $\Delta_O$ (black) and $\Delta_C$ (green), in a sample trajectory. 
The instances of $|\Delta_O(t)-\Delta_C(t)|<\delta$, which defines the transition state, are marked with red arrows. 
{\bf B}. The TSEs represented with distances of two intra-subunit residue pairs (left) and with $\alpha$ and $\beta$ angles (right).
On the two-dimensional maps representing TSE, the average positions of the TSE are marked with blue spheres, and the average distances and angles for open and closed forms are marked with the black and green circles. The calculated Tanford $\beta$-like values for the two maps are 0.33 and 0.44, respectively. 
\label{Fig_TSE}}
 \end{figure}

Allosteric regulation of protein function is encoded in the sequence space of protein family and in the native topology of protein architecture. 
Functional and energetic restraints imposed by the residue network give rise to the close correlation between  conformational fluctuations around the native state and transition dynamics in response to the ligand binding although the pattern and time scale associated with these dynamics differ in each molecule due to the stochastic nature of dynamic processes and plasticity of conformational space. 
The analysis of various inter-residue dynamics from our BD simulations revealed that upon ligand binding, the local signals at the active site of each subunit are transmitted progressively to the inter-subunit contact formation that in turn invokes the precise alignment of the catalytic elements at the active site, followed by the dimer-dimer rotation. 
Along with the experimental studies reporting the effect of mutation on the SAHH activity \cite{yuan1996JBC,ault1994JBC,vugrek2009HM}, 
our multi-faceted study of SAHH dynamics suggests that the enzymatic function is not simply represented by the relationship between the active site and ligand, but is the consequence of communication between amino acids nonlocally wired over the entire structure. 
Together with the list of evolutionarily and topologically important residues identified by the SCA and SPM (\ref{Fig_SCA}, \ref{Fig_SPM} and Table 1), the dynamic picture of $O\rightarrow C$ transitions of SAHH shown in this study is amenable to further experimental investigations.  
\\

{\bf Acknowledgement:}
This work was supported in part by the grants from the National Leading Research Lab (NLRL) program (2011- 0028885), Brain Research Center of the 21st Century Frontier Research Program (2011K000289) and the National Core Research Center (NCRC) program (2011-0006244) funded by the Ministry of Education, Science and Technology (MEST) and the National Research Foundation of Korea (NRF) (to S.C.); and the grant from World Class University (WCU) project (R31-2008-000-10010-0) (to L.S.J.); and NRF grant (2010-0000602) (to C.H.).
We thank Korea Institute for Advanced Study for providing computing resources.

\clearpage
\clearpage

\clearpage
\begin{table}[ht]
\centering
\begin{tabular}{|c||c|c|}
	\hline
	  &  cluster 1  &  cluster 2   \\
	\hline\hline
 & 19 62 65 $^{*,{\dagger}}$79 $^*$81 $^*$92 & 
81 92 \\
Perturbations  & 103 136 $^*$195 $^*$196 $^*$224 & 195 196\\
 & 293 $^*$359 $^*$386 390 392 & 293 359\\
\hline
    & 24 56 59 76 78 $^{*,{\dagger}}$79 &122 123  \\
    &  $^*$81 82 87 88 $^*$92 179&125 425  \\
Positions    & 183 192 $^*$195 $^*$196 197 208  &$^{\dagger}$426 427  \\
    &223 $^*$224 248 251 322 328  &428 429  \\
    & 349 $^*$359 368 369 $^*$386 398  &$^{\dagger}$430 431  \\
    & 406 407 420 424 & $^{\dagger}$432 \\
    \hline
\end{tabular}
\caption{Evolutionarily co-varying residues identified by statistical coupling analysis
(SCA).}
\label{Table}

\begin{flushright}$^*$ residues identified in both the perturbation and position clusters.\\
$^{\dagger}$ highly conserved residues.
\end{flushright}
\end{table}

\clearpage 

\clearpage 

\section{Supporting Figures \& Tables}
\clearpage

\renewcommand{\thefigure}{S1}
\begin{figure*}
\centering{\includegraphics[width=4in]{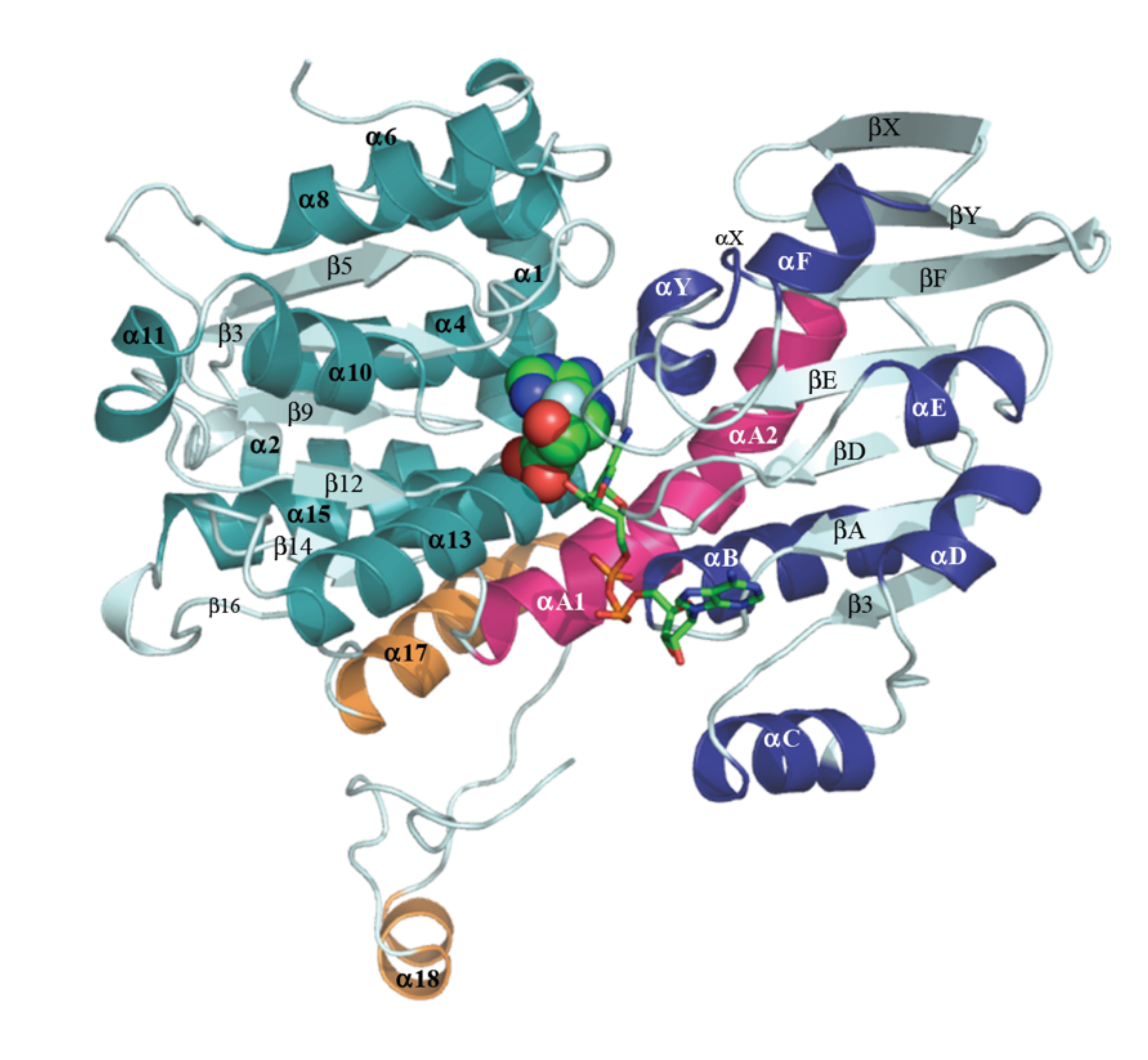}}
 \caption{The secondary structural elements in SAH hydrolase. 
The alpha helices of the catalytic, cofactor-binding and C-terminal domains are colored by cyan, blue and orange, respectively, while those of the hinge region are in magenta. The beta sheets and loops are colored by light-blue. The nomenclatures of the structural motifs follow those in the study by Turner \emph{et al.} \cite{Turner1998NSB}s: $\alpha$1, 12-28; $\alpha$2, 30-39; $\beta$3, 49-53; $\alpha$4, 58-69; $\beta$5, 73-77; $\alpha$6, 86-95; $\beta$7, 99-101; $\alpha$8, 107-118; $\beta$9, 127-130; $\alpha$10, 134-142; $\alpha$11, 144-149; $\beta$12, 152-155; $\alpha$13, 158-170; $\beta$14, 177-179; $\alpha$A1, 184-188; $\alpha$A2, 189-207; $\beta$A, 215-219; $\alpha$B, 223-234; $\beta$B, 238-242; $\alpha$C, 246-254; $\beta$C, 258-260; $\alpha$D, 262-268; $\beta$D, 271-274; $\alpha$E, 284-289; $\beta$E, 294-298; $\alpha$F, 308-314; $\beta$X, 315-322; $\beta$Y, 325-330; $\beta$F, 333-339; $\alpha$X, 340-342; $\alpha$Y, 345-349; $\alpha$15, 355-374; $\beta$16, 383-385; $\alpha$17, 388-402; $\alpha$18, 411-417. 
 \label{FigS1}}
\end{figure*}

\renewcommand{\thefigure}{S2}
\begin{figure*}
\centering{\includegraphics[width=6.2in]{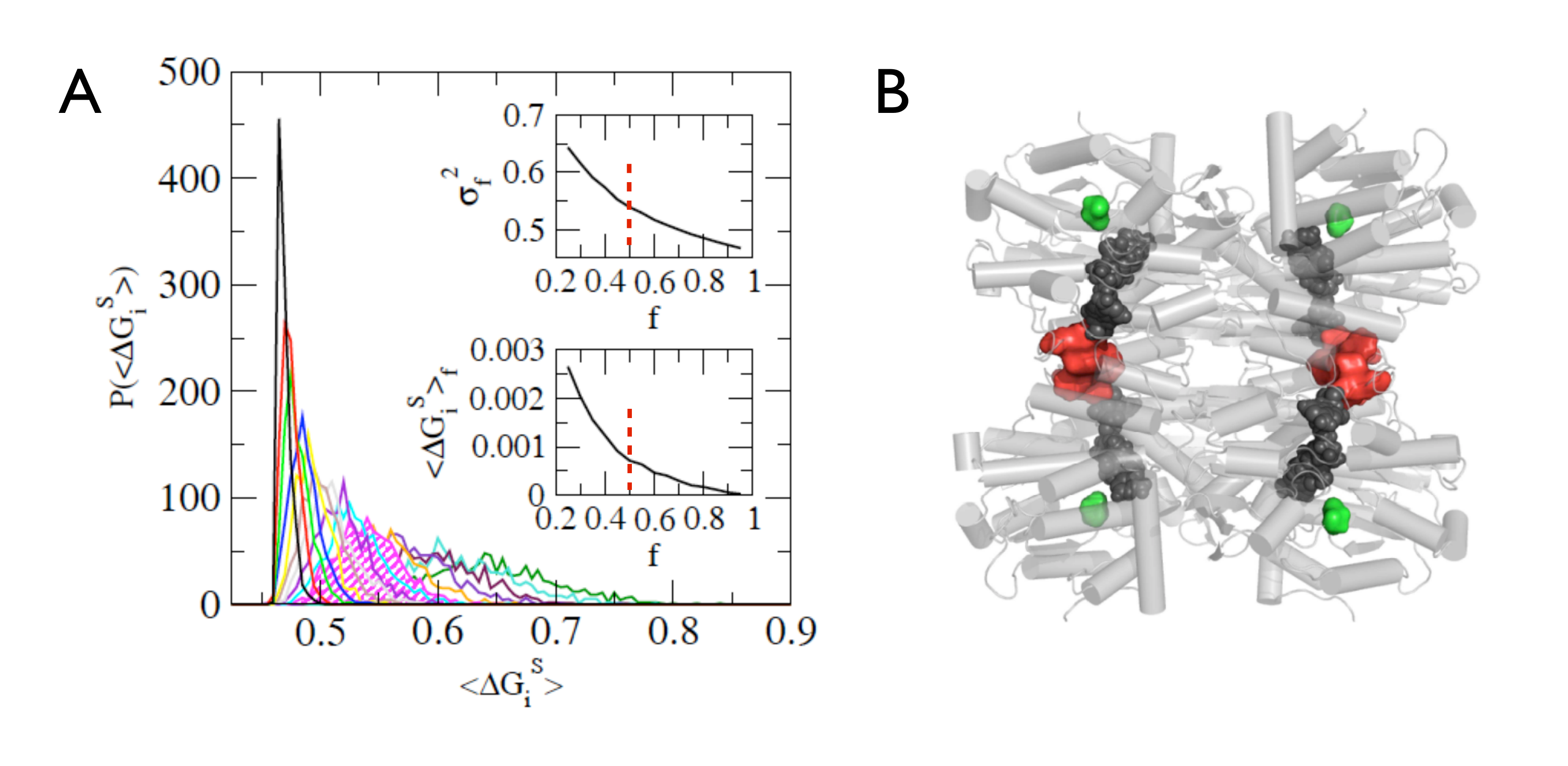}}
 \caption{Results of statistical coupling analysis.
{\bf A}. Distribution of average free energy for subalignment, $\langle G_i^s\rangle$, for various sizes ($N_s=f\times N_{MSA}$). 
{\bf B}. Among SCA-identified hot-spot residues, highly conserved residues that satisfies $\Delta G_i\geq 0.75$ are mainly found in the C-terminal domain and at C79.   
Green and red surfaces represent C79, and residues in C-terminal domain, respectively. Black spheres at each subunit represent the cofactor and ligand. 
\label{FigS2}}
\end{figure*}

\renewcommand{\thefigure}{S3}
\begin{figure*}
\centering{\includegraphics[width=6.2in]{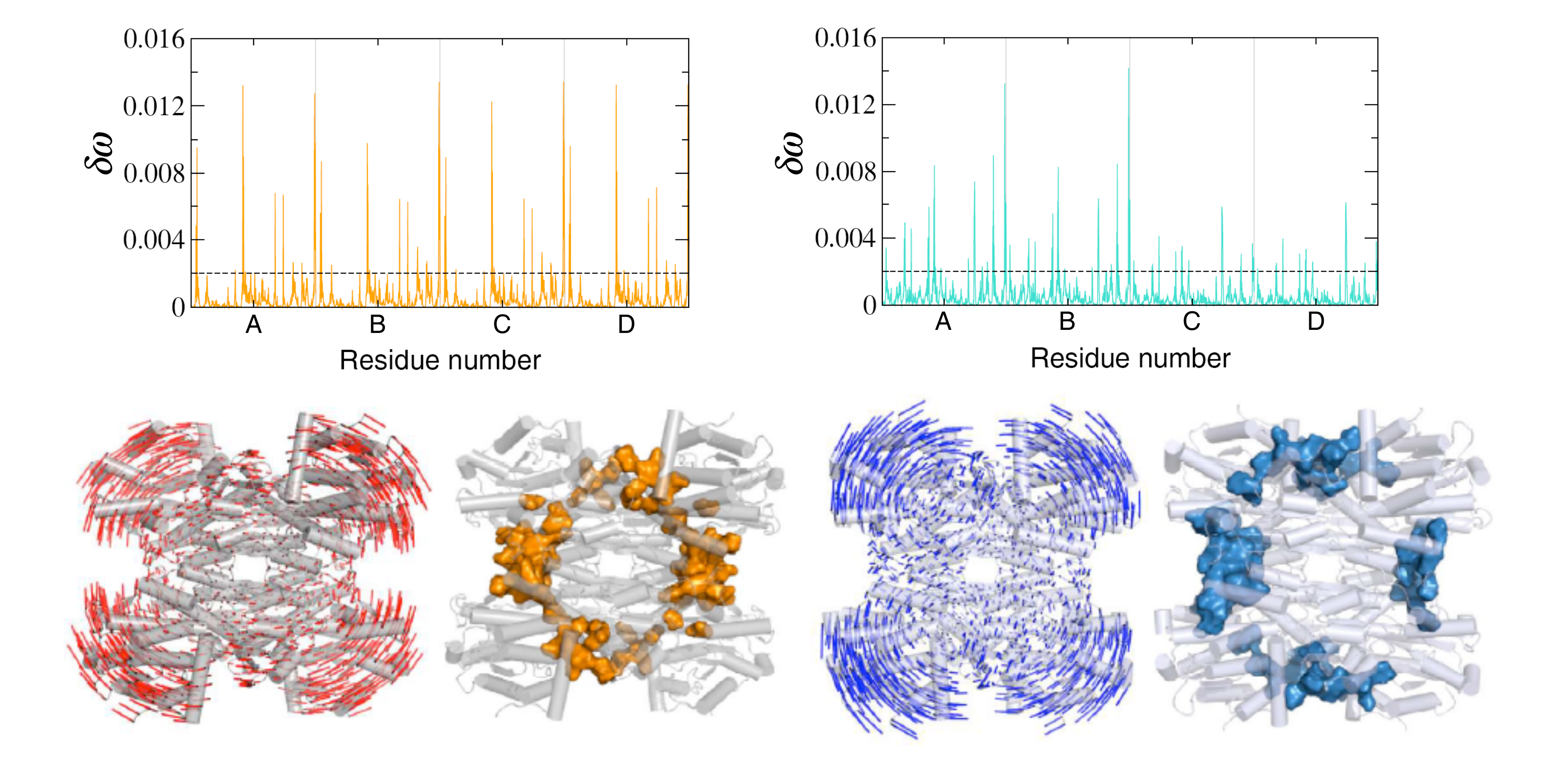}}
 \caption{Results from structure perturbation method. 
$\delta\omega$ values with respect to the next highly overlapping mode ($\vec{\nu}_{M=9}$) with $\vec{r}_{O\rightarrow C}$ for the open (orange) and closed (skyblue) forms (top). The $\vec{\nu}_{M=9}$ for the open and closed forms are depicted on the structure with red and blue lines, respectively, and listed in the Table S2. 
\label{FigS3}}
\end{figure*}

\renewcommand{\thefigure}{S4}
\begin{figure*}
\centering{\includegraphics[width=3.0in]{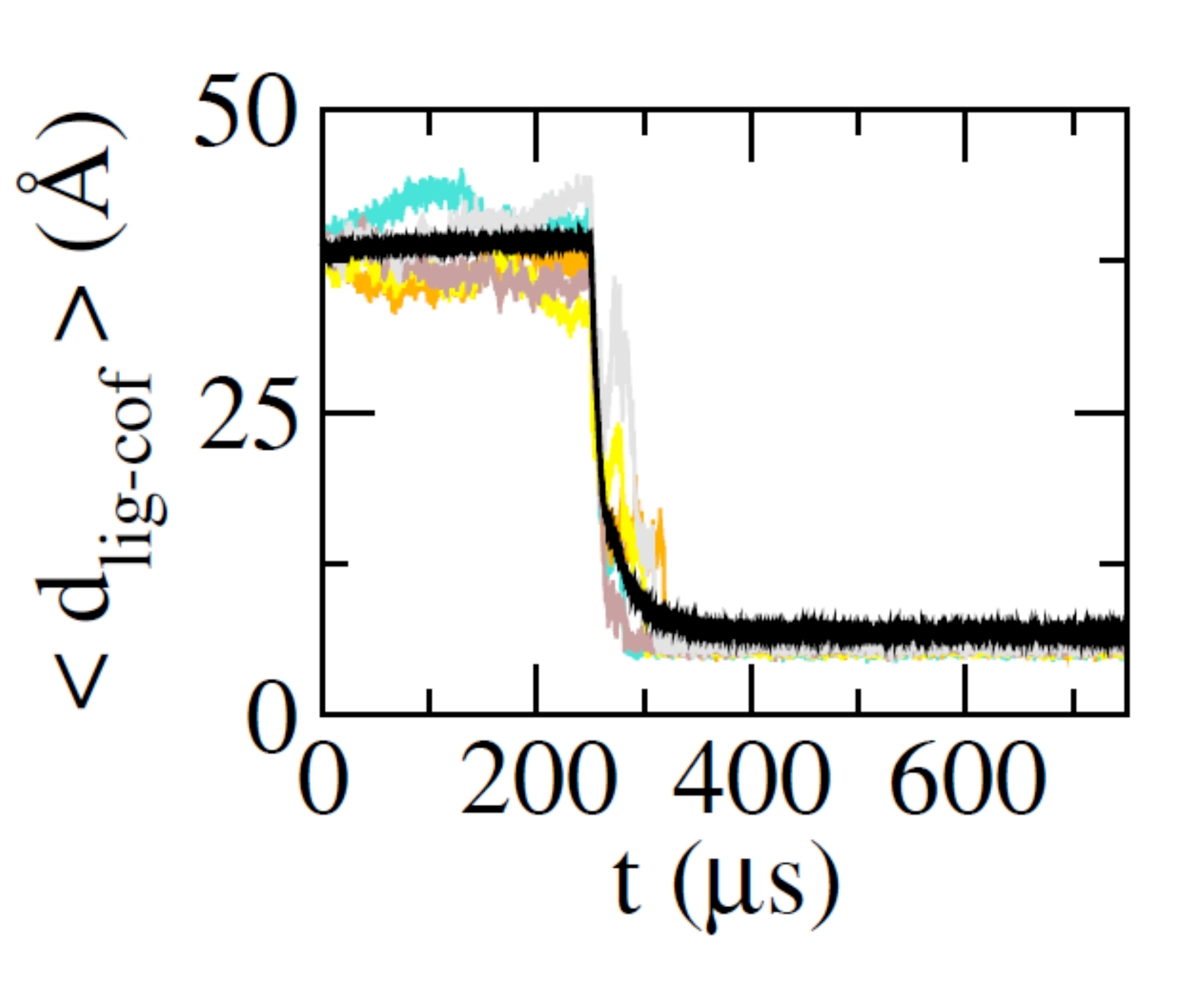}}
 \caption{ 
Ligand binding probed using the distance between the ligand and cofactor. 
\label{FigS4}}
\end{figure*}

\renewcommand{\thefigure}{Table S1}
\begin{figure*}
\centering{\includegraphics[width=5.0in]{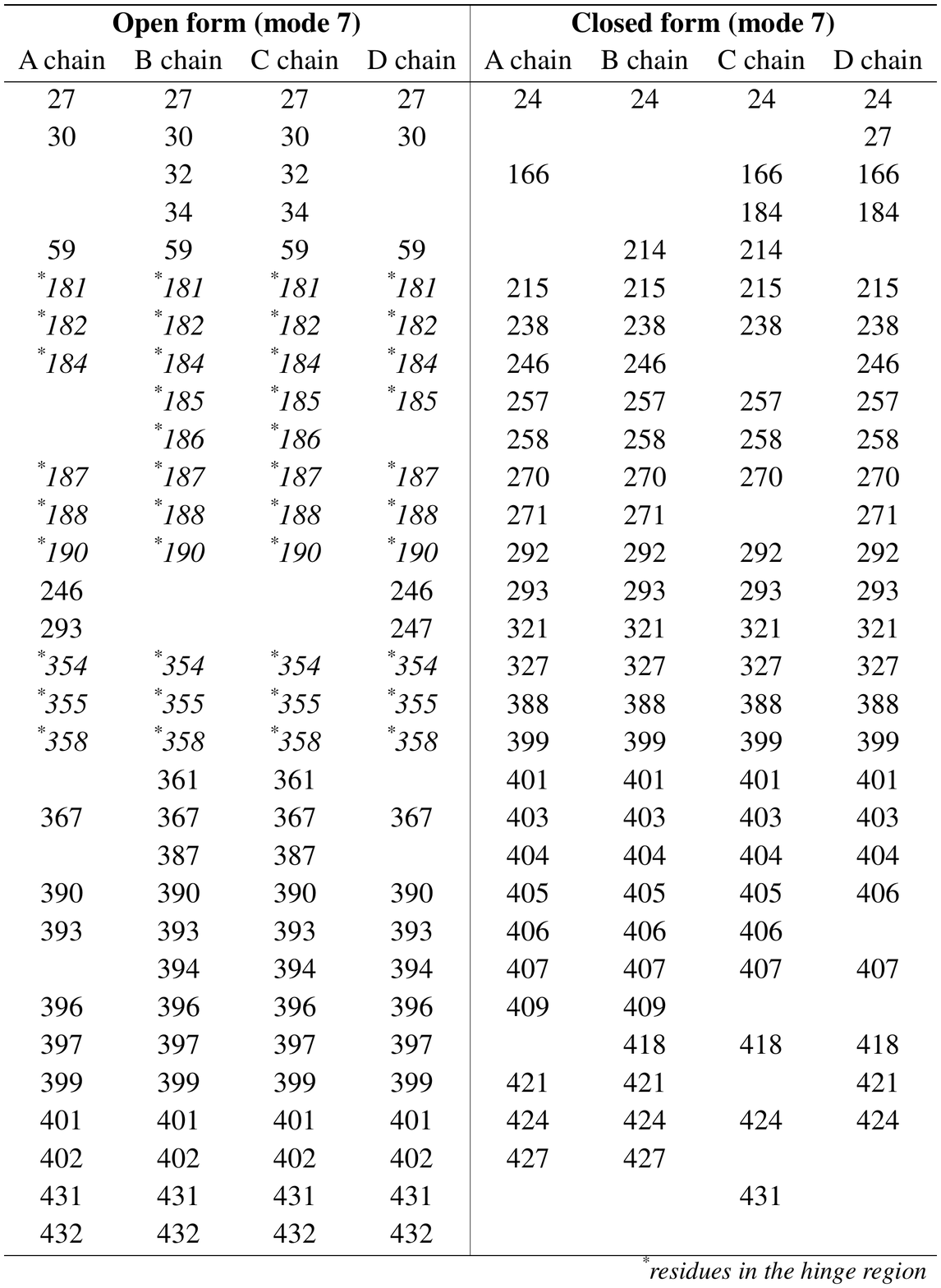}}
 \caption{ 
The list of hot spot residues identified with SPM for the mode 7 in open and closed forms.  
\label{TableS1}}
\end{figure*}

\renewcommand{\thefigure}{Table S2}
\begin{figure*}
\centering{\includegraphics[width=5.0in]{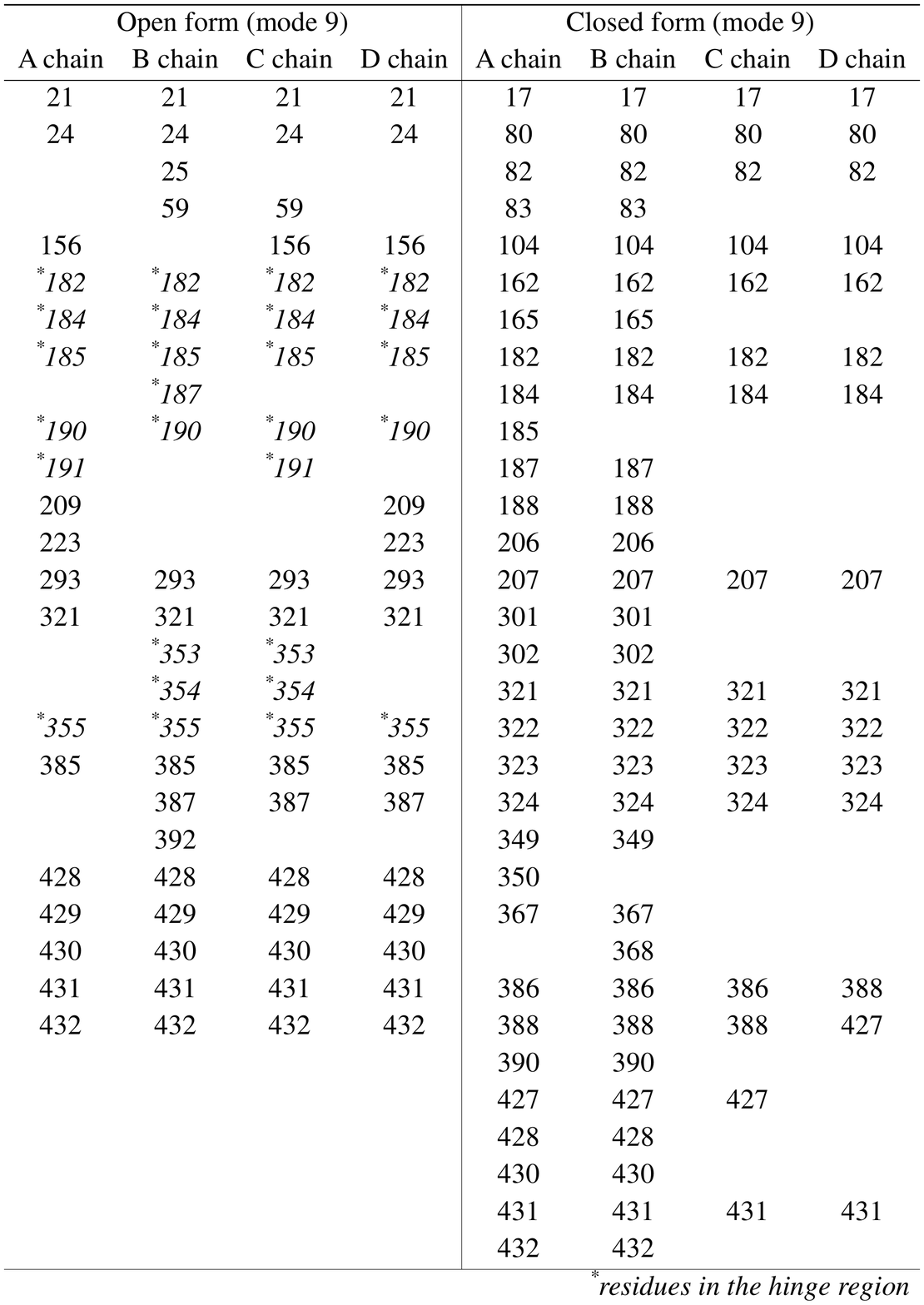}}
 \caption{ 
The list of hot spot residues identified with SPM for the mode 7 in open and closed forms.  
\label{TableS2}}
\end{figure*}

\renewcommand{\thefigure}{Movie S1}
\begin{figure*}
\centering{\includegraphics[width=5.0in]{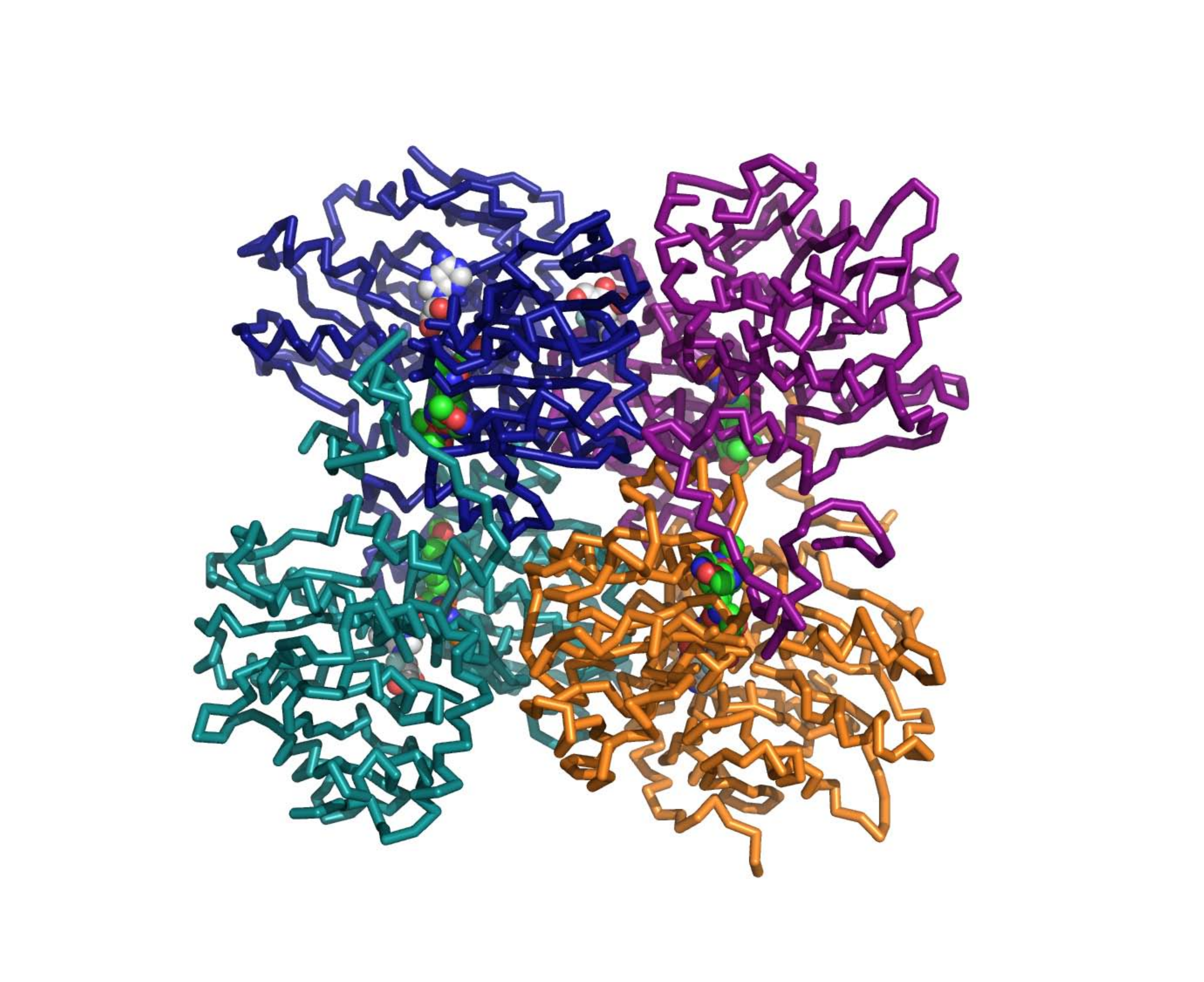}}
 \caption{A simulation of ligand-binding induced conformational change of SAHH. 
\label{movieS1}}
\end{figure*}

\renewcommand{\thefigure}{Movie S2}
\begin{figure*}
\centering{\includegraphics[width=5.0in]{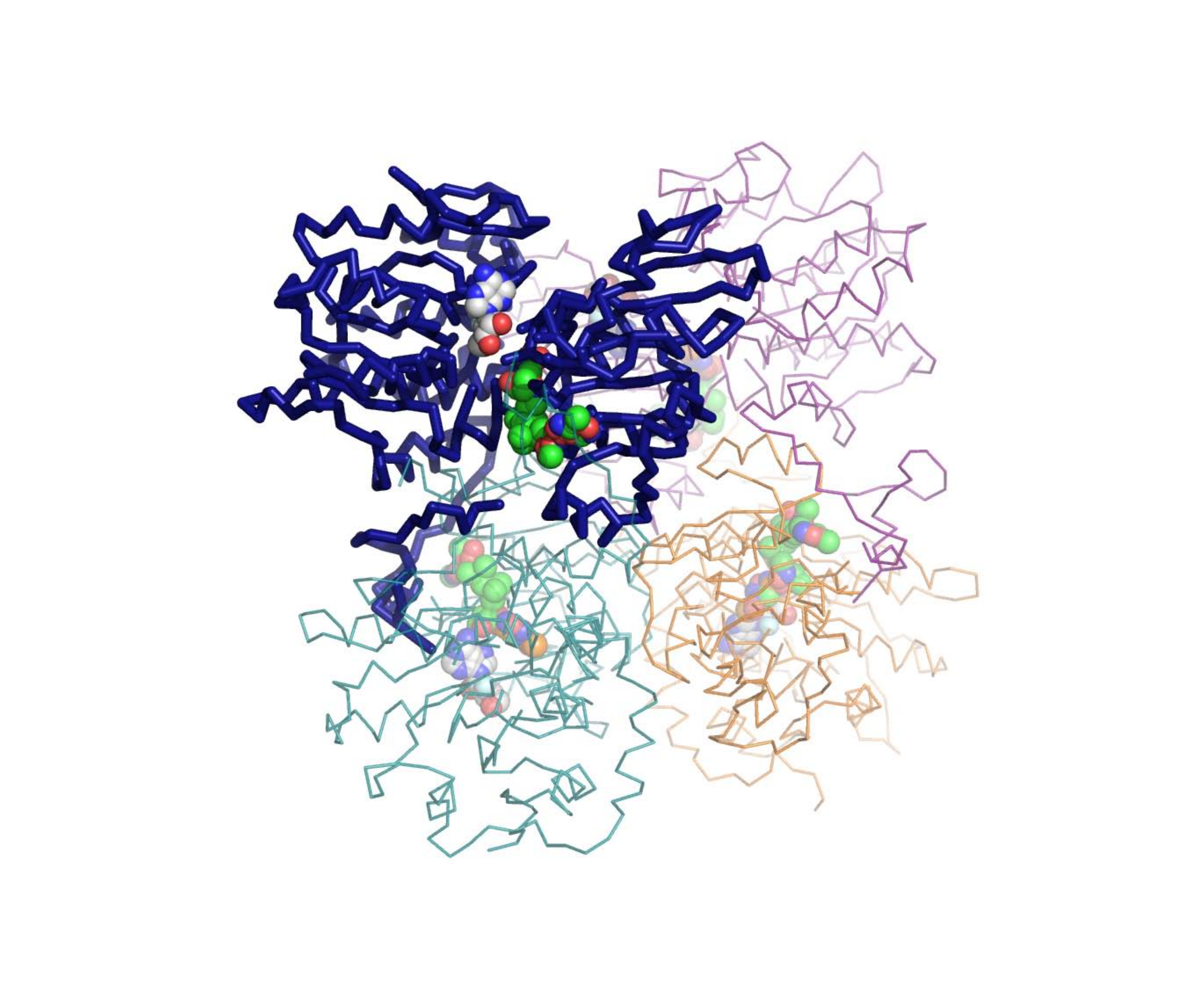}}
 \caption{A simulation of ligand-binding induced conformational change of SAHH, a cleft view 
\label{movieS2}}
\end{figure*}
\clearpage

\end{document}